\documentclass[sigconf]{acmart}
\AtBeginDocument{%
  }

\usepackage{todonotes}
\setuptodonotes{inline}
\usepackage{enumitem}
\usepackage{xspace}
\usepackage{tabularx}
\usepackage{array}
\usepackage{wrapfig}
\usepackage{capt-of}
\usepackage{float}
\usepackage{cleveref}

\setcopyright{acmlicensed}
\copyrightyear{2026}
\acmYear{2026}
\acmDOI{XXXXXXX.XXXXXXX}
\acmConference[arXiv Preprint]{}{2026}{}
\acmISBN{978-1-4503-XXXX-X/2026/11}

\newcommand{\SystemName}{\textsc{Semantic Reality}\xspace}
\newcommand{\Baseline}{\textsc{Single-Object Q\&A}\xspace}

\newcommand{\relgroup}[1]{\vspace{0.5\baselineskip}\noindent\textbf{\textsc{#1}}\ }
\newcommand{\add}[1]{{\color{black}{#1}}}

\begin{document}

\title{Semantic Reality: Interactive Context-Aware Visualization of Inter-Object Relationships in Augmented Reality}

\author{Xiaoan Liu}
\email{xiaoan@google.com}
\affiliation{%
  \institution{Google}
  \city{Seattle}
  \state{Washington}
  \country{USA}
}

\author{Eric J Gonzalez}
\email{ejgonz@google.com}
\affiliation{%
  \institution{Google}
  \city{Seattle}
  \state{Washington}
  \country{USA}
}

\author{Nels Numan}
\email{nelsn@google.com}
\affiliation{%
  \institution{Google}
  \city{Seattle}
  \state{Washington}
  \country{USA}
}

\author{Andrea Cola\c{c}o}
\email{andreacolaco@google.com}
\affiliation{%
  \institution{Google}
  \city{Mountain View}
  \state{California}
  \country{USA}
}

\author{Lucy Abramyan}
\email{abramyan@google.com}
\affiliation{%
  \institution{Google}
  \city{Mountain View}
  \state{California}
  \country{USA}
}

\author{Chen Zhu-Tian}
\email{ztchen@umn.edu}
\affiliation{%
  \institution{University of Minnesota}
  \city{Minneapolis}
  \state{Minnesota}
  \country{USA}
}

\author{Ryo Suzuki}
\email{ryo.suzuki@colorado.edu}
\affiliation{%
  \institution{University of Colorado Boulder}
  \city{Boulder}
  \state{Colorado}
  \country{USA}
}

\author{Mar Gonzalez-Franco}
\email{margon@google.com}
\affiliation{%
  \institution{Google}
  \city{Seattle}
  \state{Washington}
  \country{USA}
}

\renewcommand{\shortauthors}{Liu et al.}

\begin{abstract}
Bridging the physical and digital world through interaction remains a core challenge in augmented reality (AR). Existing systems target single objects, limiting support for planning, comparison, and assembly tasks that depend on relationships among multiple items. We present \SystemName, an AR system focused on surfacing inter-object connectivity and making it interactive. Leveraging multimodal reasoning, spatial anchoring, and physical action recognition, \SystemName maintains a persistent model of objects around the user and their relationships. Connections are visualized in-situ to highlight compatibility, reveal next steps, and reduce ambiguity during tasks. We contribute a connectivity-centered interaction paradigm and a system architecture that couples anchor tracking, action sensing, and model inference to construct a live connectivity graph. In an exploratory study comparing \SystemName to a single-object baseline, participants reported clearer inter-object understanding and higher engagement and satisfaction, without increased perceived task load. A scenario study illustrates where connectivity aids planning, sequencing, and disambiguation.
\end{abstract}



\begin{teaserfigure}
  \includegraphics[width=\textwidth]{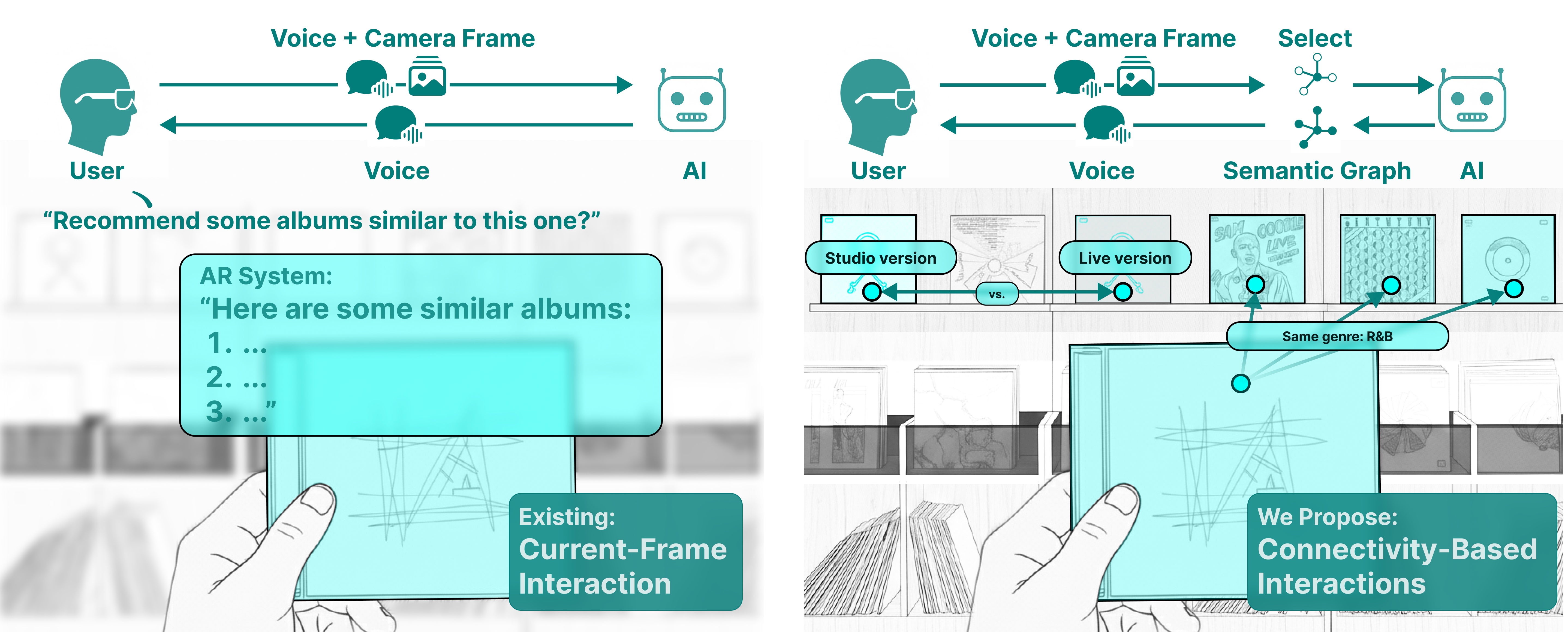}
  \caption{Existing current-frame interaction (left) treats the camera view as a single query and overlays a text answer. Our approach (right) adds selection that writes into a scene-anchored semantic graph, making relations between objects visible and actionable. The graph supports connectivity-based interactions such as grouping, comparison, and compatibility checks, while keeping guidance grounded in the physical scene.}
  \Description{Side-by-side teaser. Left: user asks for similar albums and receives a text list from a current-frame query. Right: user selection creates a semantic graph that connects multiple albums with labeled relations for connectivity-based interactions.}
  \label{fig:teaser}
\end{teaserfigure}


\maketitle

\section{Introduction}
Augmented reality (AR) and multimodal large language models (MLLMs) are beginning to meet in everyday settings \cite{suzuki6programmable}. With access to egocentric cameras and scene tracking, MLLMs can recognize objects and provide grounded answers about what the user sees~\cite{suzukiEverydayARAIintheLoop2024, bovoEmBARDimentEmbodiedAI}. By blending multimodal input and spatially anchored output overlaid on the real world, AR+AI promises interactions that move beyond a simple text exchange. Early examples of the potential of this blend have already been explored in projects like XR-Objects \cite{doganAugmentedObjectIntelligence2024} or RealityProxy \cite{liu2025reality}. However, realizing real-time, world-grounded collaboration with an AI, is far beyond simple information retrieval, and it certainly requires interface abstractions that go beyond making model outputs precise. We propose that such grounded AI collaboration will have to allow for new levels of semantic understanding that can be referential and help establish an actionable plan among the objects in the scene. 

Progress toward physically grounded Human–AI symbiosis~\cite{bovo2025symbioticaiaugmentinghuman} requires the physical world to become an active participant rather than a passive canvas. Modern XR platforms blend virtual content with physical surroundings through scene understanding, yet still treat the world as a backdrop for displays rather than a center of reasoning and action~\cite{Ruofei2020depthlab, DBLP:conf/chi/NuernbergerOBW16, lindlbauer2018remixed}. Recent work on Augmented Object Intelligence (AOI) equips everyday objects with interactive digital capabilities, blurring boundaries between physical artifacts and computational services~\cite{doganAugmentedObjectIntelligence2024}. Building on this trajectory, we ask how AR systems can move beyond recognizing individual objects to representing and operating on the \emph{relationships} (what we call \emph{connectivity}) that tie multiple items together in real tasks.

Many real-world activities hinge on such relationships: planning the order of steps, checking whether parts are compatible, comparing alternatives, or understanding what tool acts on which component. Yet most AR+AI interactions remain centered on single objects and free-form text responses. Without an explicit, scene-anchored representation of connections, users shoulder heavy cognitive indexing across items, while the AI must negotiate ambiguous references and cannot easily point back to specific entities in the world~\cite{Ruofei2020depthlab, DBLP:conf/chi/NuernbergerOBW16, lindlbauer2018remixed, gonzalez2024guidelines}.

To address this, we present \SystemName, a connectivity-centered substrate \cite{mackay2025interaction} between perception and inference. \SystemName maintains a dynamic \emph{semantic graph} of the surrounding scene in which nodes are detected physical objects and edges are typed inter-object relations.
This shared scaffold lets the AI refer to concrete entities and links, and lets the AR runtime render unambiguous, in-situ overlays grounded in the user's environment rather than abstract text. The representation evolves as users select or manipulate objects and as task context changes, localizing familiar knowledge to the specific items at hand. 


We articulate the underlying concepts in a design space organized around three aspects.
First, \textbf{\emph{relation types}} capture the types of connections our system makes actionable. We focus on eight types: four descriptive (i.e., \emph{spatial}, \emph{structural}, \emph{similarity}, \emph{comparison}) and four prescriptive (i.e., \emph{affordance}, \emph{compatibility}, \emph{procedural}, \emph{causality}). 
Second, the \textbf{\emph{interaction initiative}} distinguishes \emph{user-initiated}, \emph{system-suggested}, and \emph{hybrid flows}, clarifying where a relation originates and who holds agency at a given moment. 
Third, a scene-anchored \textbf{\emph{context window}} defines the active subgraph in play, including a user node whose edges record interactions such as holding, pointing, and proximity. This framing constrains inference and guides presentation without requiring heavy authoring. 

We operationalize these ideas in a prototype running on a head-mounted display. The system continuously detects objects and anchors them in world coordinates; it then infers typed relations within the current context window using MLLM reasoning constrained to the eight relation types. Accepted edges are rendered as consistent, in-situ overlays. Users steer and refine the graph through gaze-and-pinch multi-selection, voice requests, and embodied gestures such as grabbing and bringing objects together. Our approach builds on AOI~\cite{doganAugmentedObjectIntelligence2024} and related AR graph methods, but shifts the focus from single-object lookup to interactive connectivity.

We evaluate \SystemName in two parts.
First, an exploratory, within-subjects study compares \SystemName to a single-object baseline adapted to our hardware. Participants reported clearer inter-object understanding and higher engagement and satisfaction without increased perceived task load.
Second, a complementary benchmarking study uses short scenario videos to contrast \SystemName with common alternatives (unaided real world, YouTube, chat assistant, and a single-object XR baseline), characterizing when connectivity aids planning, disambiguation, and safety.

The contributions of this work are:
\begin{itemize}
  \item A connectivity-centered substrate for AR+AI, formalized as a dynamic, scene-anchored semantic graph of inter-object relations, and a design space that organizes relation types, interaction initiative, and context window.
  \item A system that operationalizes connectivity by coupling open-vocabulary object detection, world anchoring, and constrained MLLM reasoning to infer typed relations and render them in situ.
  \item Interaction techniques that let users steer and confirm connectivity through multi-selection, voice, and embodied gestures, with real-time updates to the semantic graph.
  \item Two evaluations comparing the connectivity-centered approach to common alternatives, indicating improved relation clarity and engagement without added task load, and identifying scenarios where connectivity provides the most benefit.
\end{itemize}






\section{Related Work}
\SystemName{} integrates visual input and object segmentation to maintain global context of scene entities for consistent, context-aware responses. We review previous literature on Physical AR Interfaces, Contextually Adaptive Interfaces, and Reasoning of Physical Environments.

\subsection{Physical AR Interfaces}

Seamless integration of physical and virtual worlds requires environmental understanding. \textit{Light Anchors}~\cite{ahuja2019lightanchors} spatially anchors virtual elements using point lights, Reality Editor~\cite{heun2013reality} visualizes relationships among smart objects, and \textit{RealityCheck}~\cite{hartmann2019realitycheck} and \textit{FLARE}~\cite{gal2014flare} leverage context for immersive AR. Early approaches explored leveraging physical affordances within arm's reach~\cite{suzuki2021hapticbots}, using everyday items as tangible proxies~\cite{du2022opportunistic} that could be re-rendered~\cite{hettiarachchiAnnexingRealityEnabling2016}. PapARVis~\cite{papARvis, tong2022exploring} uses paper tangibility to interact with data visualizations through direct manipulation.


Recent approaches enable physical objects to act as portals to digital information through AI. Most relevant, \textit{XR-Objects}~\cite{doganAugmentedObjectIntelligence2024} employs real-time object segmentation combined with MLLMs for context-aware interactions without manual pre-registration, aligning with \textit{Reality Summary}~\cite{gunturu2024realitysummary} in enhancing the semantic depth of AR interactions. These systems illustrate a shift towards transforming everyday objects into interactive digital entities. We build upon this by enabling richer interactions across scales, from within-object details to inter-object relationships.

\add{While many systems focus on individual objects, earlier AR research considered inter-object relationships, such as spatial relationship graphs for modeling dependencies between tracked objects~\cite{echtler2008splitting} and semantic reasoning over object-to-object relations~\cite{semantic_jensen_2009}. These established the value of connectivity in AR. Our work builds on this but shifts from primarily spatial or pre-defined links to an interactive, user-steered model. \SystemName infers eight typed, task-oriented relations and makes the resulting graph a primary medium for interaction rather than a background data structure.}

\subsection{Contextually Adaptive Interfaces in AR}
Contextual adaptation is central in AR because decisions about where and how to display elements must respond dynamically to the user's context~\cite{DBLP:journals/tvcg/GrubertLZR17}.
Much work achieves adaptability through geometric properties: aligning elements with physical edges~\cite{DBLP:conf/chi/NuernbergerOBW16}, anchoring to planar surfaces~\cite{DBLP:journals/tvcg/ChenS0WQW20}, or integrating with 3D meshes~\cite{DBLP:conf/uist/FenderLHA017, DBLP:conf/chi/FenderHA018}.
BlendMR~\cite{hanBlendMRComputationalMethod2023} optimizes for visual clarity and geometry-fit when blending MR content onto surfaces.
AdapTUI~\cite{heAdapTUIAdaptationGeometricFeatureBased2024} enables tangible UIs to adapt via geometric feature recognition.

Beyond geometry, systems incorporate semantic information to guide adaptation.
Lindlbauer et al.~\cite{lindlbauerContextAwareOnlineAdaptation2019} regulate detail and positioning based on context and cognitive load.
Tahara et al.~\cite{DBLP:conf/ismar/TaharaSNI20} use scene graphs for consistent adaptation across locations.
AdapTutAR~\cite{DBLP:conf/chi/HuangQWPSCRQ21} adjusts instructional content based on user characteristics,
while SemanticAdapt~\cite{chengSemanticAdaptOptimizationbasedAdaptation2021} associates object categories with relevant overlays. Zhu-Tian et al.~\cite{DBLP:journals/tvcg/ChenCLYBP24} use reinforcement learning to adapt AR label layout in changing environments.
AdjustAR~\cite{numanAdjustARAIDrivenInSitu2025} applies MLLMs to correct semantic misalignments between site-specific AR content and a physical environment that has changed since authoring time.

Unlike these pre-defined adaptation approaches, our system leverages AI to automatically generate and adapt interfaces based on real-time context, similar to Reality Proxy~\cite{liu2025reality}.

\add{A related challenge is view management---avoiding visual clutter and label overlap as complexity increases. Foundational work established annotation layout techniques using constraints and optimization~\cite{bell2001view, kalkofen2008comprehensible, grasset2012image}. Our prototype employs basic heuristics for placing overlays; dense scenes with many relations would require more sophisticated strategies such as automatic label placement and decluttering. Our contribution focuses on the semantic graph itself, complementing established view management principles.}


\subsection{Perception and Reasoning in Physical Environments}

AR interaction demands effective methods for real-time scene understanding. Techniques for semantic segmentation and MLLMs have significantly enhanced context awareness over earlier approaches relying on geometric features~\cite{galFLAREFastLayout2014} or manually defined labels~\cite{chenPaperToPlaceTransformingInstruction2023}.

Deep learning models such as YOLO~\cite{redmonYOLOv3IncrementalImprovement2018} and Mask R-CNN~\cite{heMaskRCNN2018} support reliable recognition of predefined categories and have been applied in constrained HCI settings. For example, CookAR~\cite{leeCookARAffordanceAugmentations2024} augments kitchen tools for low-vision users by segmenting functional parts, and Lang et al.~\cite{langVirtualAgentPositioning2019} use segmentation to guide virtual agent positioning.

Recent MLLMs support open-vocabulary understanding and flexible multimodal interpretation. Models such as GPT-4V~\cite{GPT4VisionSystemCard2023}, Gemini~\cite{team2023gemini}, and LLaVA~\cite{liuVisualInstructionTuning2023} support spatial reasoning, grounded dialogue, attribute comparison, and action recognition~\cite{liuMMBenchYourMultimodal2024}. Human-centered systems leverage these capabilities: Human I/O~\cite{liuHumanUnifiedApproach2024} combines egocentric sensing with MLLM prompting for situational impairment detection; SpaceBlender~\cite{numanSpaceBlenderCreatingContextRich2024} reasons about unseen regions; and XaiR~\cite{srinidhiXaiRXRPlatform2024} guides multistep physical tasks.

\section{Semantic Reality}
\label{sec:concept}

\begin{figure*}[h]
    \centering
    \includegraphics[width=\linewidth]{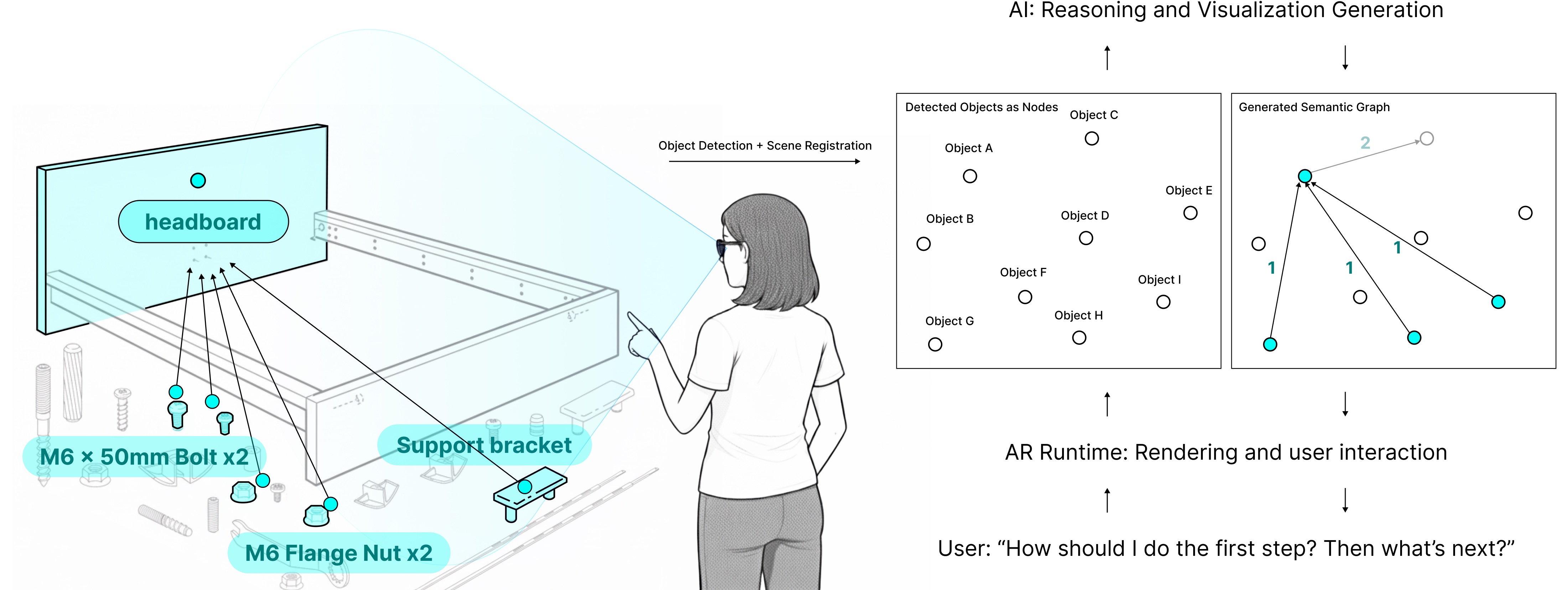}
    \caption{Concept overview of \SystemName. Left: in a furniture assembly scene, the AR overlay labels detected parts and projects connections in situ. Right: detected objects become nodes; the system infers typed edges that the AR runtime maps back to the scene.}
    \label{fig:concept_overview}
\end{figure*}

\SystemName is a connectivity-centered \emph{substrate} between the AR runtime and the AI layer. It maintains an explicit, scene-anchored semantic graph of how real-world objects relate to one another, updated continuously as the system detects objects, users make selections, and task context evolves (\autoref{fig:concept_overview}).

\subsection{Semantic Graph Substrate}
\label{subsec:scenegraph}
The substrate is a typed, attributed graph aligned to world coordinates. \textbf{Nodes} represent scene-anchored physical objects, each storing a stable identifier, a canonical label, and geometric metadata (3D pose and approximate extent). \textbf{Edges} encode inter-object relations as typed links corresponding to the eight relation types below; each edge carries lightweight attributes such as confidence and is generated on demand as context changes. This representation is compatible with prior AR scene graph approaches~\cite{DBLP:conf/ismar/TaharaSNI20,doganAugmentedObjectIntelligence2024,gunturu2024realitysummary}. As the graph is scene-indexed, the AI can target concrete nodes and edges when formulating guidance, and the AR runtime can place overlays at the corresponding physical locations. The graph serves three complementary roles: \emph{referential precision}, allowing AI outputs to target concrete nodes and edges rather than generic labels; \emph{shared memory}, accumulating user-confirmed relations so that subsequent guidance builds on what has been established; and \emph{scene-grounded semantics}, situating familiar knowledge about how things work in the current environment, filtered by the user's selections and task constraints.

\begin{figure*}[h]
    \centering
    \includegraphics[width=0.80\linewidth]{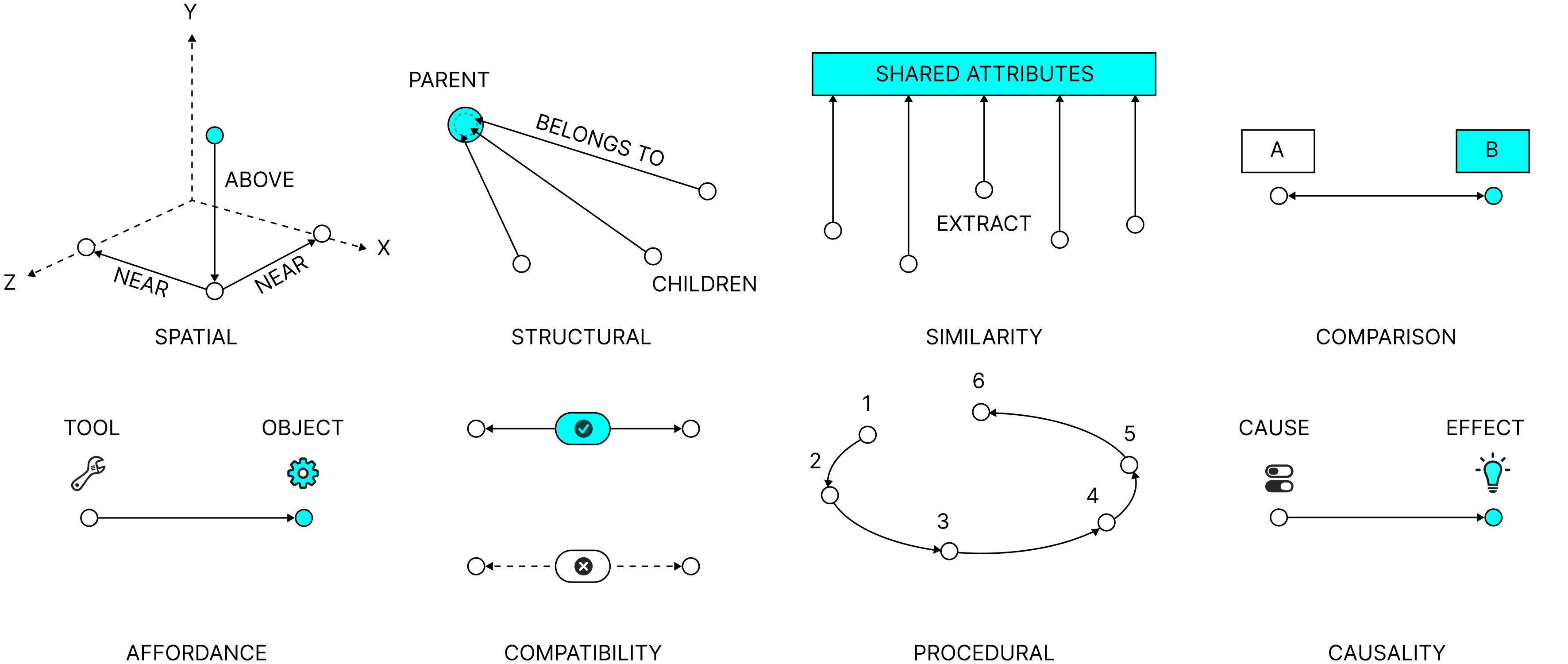}
    \caption{Overview of the eight relation types used by \SystemName. Top row: spatial, structural, similarity, comparison. Bottom row: affordance, compatibility, procedural, causality. Encodings are consistent across the UI to help users quickly read links in context.}
    \label{fig:rel_banner}
\end{figure*}
\vspace{-2mm}

\label{subsec:designspace}
We further articulate the design space of connectivity along the eight relation types identified below. This framing constrains inference and guides how the AR system presents connections.

\add{
\subsection{Deriving the Relation Taxonomy}
To ensure our design space covers the breadth of real-world physical interactions, we adopted an exemplar-driven approach. Our process followed three stages: scenario collection, abstraction and induction, and iterative refinement.

\subsubsection{Scenario Collection}
We gathered a corpus of 52 diverse interaction scenarios (detailed in Supplemental Material). We recorded 28 task sessions in our lab spanning cooking, electronics assembly, and shelf organization, and collected 24 clips from DIY and instructional channels on YouTube. This dataset captured a wide range of user intents, from identifying parts in a makerspace to comparing products in a grocery aisle.

\subsubsection{Abstraction and Induction}
We performed an inductive analysis to identify the implicit connections users relied on. We coded each scenario by the ``linking verbs'' that connected objects, such as ``fits into'' and ``works with'' (which converged into \textit{Compatibility}), ``looks like'' and ``same kind as'' (\textit{Similarity}), and ``must be done before'' versus ``enables'' (which split into \textit{Procedural} and \textit{Causality}, respectively). The resulting descriptive--prescriptive grouping aligns with established categorizations of relational knowledge in cognitive science, which distinguish taxonomic and spatial relations from functional and causal ones~\cite{gentner1983structure}.

\subsubsection{Iterative Refinement}
Two of the authors refined these candidates through a series of iterative coding sessions. Disagreements were resolved through discussion to establish mutually exclusive categories. For instance, we distinguished \textit{affordance} (a potential action) from \textit{compatibility} (a constraint check). This process consolidated the candidates into the final eight relation types, organized into descriptive and prescriptive categories.}

\subsection{Relation Types}
We focus on eight types that proved actionable in our prototype (\autoref{fig:rel_banner}; illustrative examples in Appendix~\ref{appendix:relation_figures}). We group them into \emph{descriptive} types that characterize the current scene and \emph{prescriptive} types that suggest how to act.

\relgroup{Descriptive relations.}
\textbf{Spatial} relations capture topological or metric configuration (\emph{on}, \emph{in}, \emph{near}). They are sensed from geometry and tracking and support wayfinding and disambiguation, e.g., localizing keys by their position relative to nearby books.
\textbf{Structural} (part--whole) relations describe how objects compose larger entities (e.g., screw \emph{part-of} bracket), providing stable constraints that collapse tool or part candidates into those belonging to a selected assembly.
\textbf{Similarity} captures shared attributes or roles across a set of objects (e.g., books on the same topic), enabling zoom-out summaries and quick clustering.
\textbf{Comparison} foregrounds differences between alternatives (e.g., contrasting two shampoos on ingredients and price), supporting side-by-side decision making.

\relgroup{Prescriptive relations.}
\textbf{Affordance} links express that one entity can act upon another (e.g., knife \emph{acts-on} garlic), surfacing viable actions with concise technique tips.
\textbf{Compatibility} encodes whether entities fit or function together (e.g., USB-C plug \emph{fits-with} port), including parameter constraints, warnings, and substitution suggestions.
\textbf{Procedural} links capture step ordering and parallelism opportunities. The system generates lightweight plans with numbered steps that update as users progress.
\textbf{Causality} records preconditions and effects (e.g., drilling a pilot hole \emph{enables} fastening), supporting what-if predictions and justifying recommendations.

\section{Implementing \SystemName}
This section describes how \SystemName infers inter-object relations, renders them as AR overlays, and lets users update the graph through selection, voice, and embodied gestures.

\subsection{Inferring Connectivity}
\label{sec:impl-infer}

\begin{figure*}[h]
    \centering
    \includegraphics[width=\linewidth]{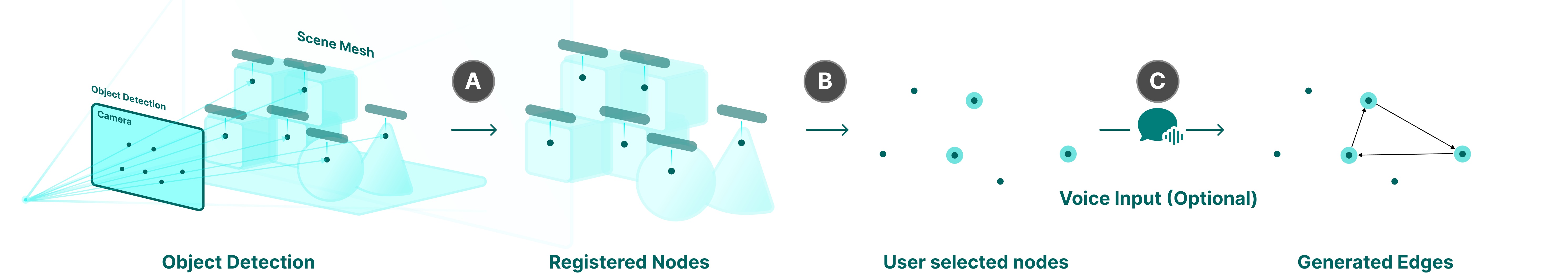}
    \caption{Inference pipeline in \SystemName. Object detections are anchored to the scene mesh to create registered nodes \textbf{(A)}. Users nominate a subset through selection to form the active reasoning context \textbf{(B)}. An optional voice request specifies the desired operation or constraints \textbf{(C)}. Conditioned on this context and request, the system proposes typed edges among relevant nodes and updates the semantic graph.}
    \label{fig:infer_pipeline}
\end{figure*}

\noindent
\autoref{fig:infer_pipeline} summarizes the end-to-end flow used to infer connectivity. Detections are first registered as scene-anchored nodes, then user selections establish the active context for reasoning. Optional voice input specifies the intended operation, after which constrained inference proposes candidate edges and commits accepted links to the graph.

\subsubsection{Maintaining the semantic graph}
\SystemName maintains a dynamic semantic graph whose nodes correspond to detected physical objects and whose edges capture typed relations between them. Nodes are instantiated by recurring perception on the Apple Vision Pro: we capture passthrough frames and perform \emph{open‑vocabulary 2D detections} using a multimodal large model (Gemini 2.5 Flash) to obtain regions and semantic labels for likely objects \cite{gemini25flash}. Each detection is associated with the current camera pose and projected into world space by \emph{raycasting} through the 2D region into the live scene mesh provided by RealityKit scene reconstruction, yielding a 3D anchor in world coordinates \cite{appleRealityKit}. For each detected region, we also persist a cropped RGB patch to support later re‑identification and attribute reasoning.

We implement the AR runtime in Unity targeting visionOS, using PolySpatial to bridge world‑anchored Unity content to RealityKit on device \cite{unityPolySpatial,appleRealityKit}. Each node stores a stable identifier tied to its anchor, current 6‑DoF pose, and semantic attributes (canonical label, synonyms). When subsequent frames refine pose or label, we update the node in place rather than duplicating it. This yields a persistent, scene‑anchored set of nodes that higher layers can reference reliably, as shown in \autoref{fig:infer_pipeline}A.

\subsubsection{Context windows}
At any moment, only a subset of nodes is relevant. \SystemName forms an \emph{active reasoning context} (or context window) that combines (i) user-selected nodes, (ii) recently manipulated or proximate nodes, and (iii) nodes mentioned in the user's request (\autoref{fig:infer_pipeline}B). This context induces a focused subgraph on which relation inference is run, preventing spurious relations in cluttered scenes. The window evolves as users look, point, grasp, or multi-select, narrowing attention to currently relevant relations.

A distinguishing feature of the graph is that the \emph{user is represented as a node}. Edges from this node capture embodied interactions such as holding, pointing, and arranging, inferred from hand tracking and VLM reasoning. Spatial proximity further modulates the window, giving higher weight to items within reach or field of view. This treatment makes the graph explicitly agent-centric: the context window is always anchored to the user's body and attention, supporting rapid, local updates when focus shifts.

\add{
\subsection{Underlying Inference Mechanism}
To robustly infer relationships, the system employs a staged reasoning process that decouples intent recognition from detailed attribute extraction. This separation prevents the model from conflating distinct relation types and ensures adherence to the strict schema required for visualization.

\subsubsection{Intent Classification}
First, the system determines the intended relation type based on the user's action, selected objects, and any accompanying voice input. For example, when a user selects two items (with or without an explicit command like ``Compare these''), the system triggers a classification prompt (see \autoref{appendix:prompts}) to decide if the interaction implies a comparison, compatibility check, or structural connection. This step acts as a router, narrowing the search space to a single relation category.

\subsubsection{Parameter Extraction}
Once the relation type is established, the system executes a specialized prompt designed solely for that type. This prompt retrieves the specific edge attributes necessary for rendering, such as identifying the ``parent'' and ``child'' nodes for a structural link or extracting distinct feature differences for a comparison. By isolating this task, we ensure high-fidelity outputs grounded in the current scene context.

\subsubsection{Ambiguity Resolution}
In cases where multiple relation types are plausible for the same selection (e.g., two devices could be both \emph{compared} or checked for \emph{compatibility}), the system detects this ambiguity during the classification stage. Rather than guessing, it elicits user intent via a disambiguation prompt (e.g., ``Compare these two chargers, or show compatibility?''). User confirmation resolves the choice and commits the corresponding edge(s) into the graph, keeping the model's proposals aligned with user goals.
}

\subsection{Visualization of the \SystemName}
\label{sec:impl-present}

Our prototype runs on Apple Vision Pro using Unity, with PolySpatial to render world-anchored content via RealityKit on visionOS~\cite{unityPolySpatial,appleRealityKit}. Once the system commits a graph update, the front end instantiates corresponding primitives: anchored billboards for labels, world-space polylines for connectors, and small cards for edge-specific details. All overlays are depth-tested against the scene mesh to respect occlusion. Edges are ephemeral by default and decay when they no longer match the current context.

As the user interacts, inference produces deltas that update overlays incrementally. Typed edges map to consistent visual encodings: structural relations appear as lightweight labels near parts; comparisons as midline cards with key attributes; compatibility and affordance links as directional connectors with captions; procedural and causal links with step numbers or ``enables'' arrowheads; and spatial relations as minimal depth-respecting ribbons.

\add{We discuss potential risks and implications of visual clutter in \cref{sec:visual_complexity}, as well as relevant improvements that could be explored in future work.}

\subsection{Interacting with \SystemName}

Connectivity links can be \emph{user-initiated} (e.g., grasping a screwdriver and pointing at a screw triggers inference for that subset), \emph{system-initiated} (e.g., pre-highlighting compatible ports near a cable), or \emph{hybrid} (a brief selection seeds the substrate, then the system expands with related edges). User-initiated relations are visualized with stronger emphasis reflecting explicit intent, while system-initiated ones are presented tentatively behind lightweight confirmations to manage clutter.

\label{sec:impl-interact}
\begin{figure}[t]
    \centering
    \includegraphics[width=\linewidth]{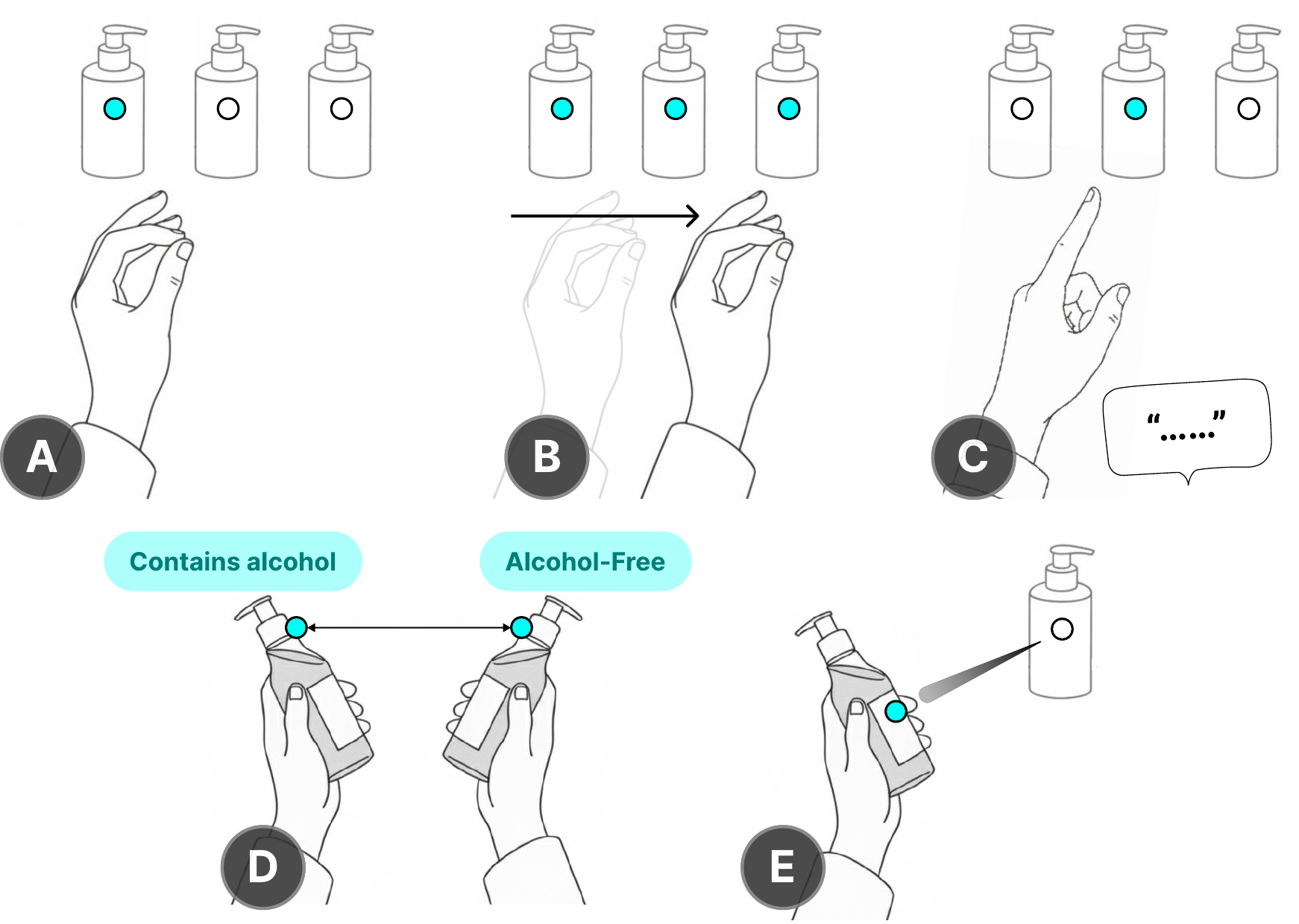}
    \caption{Interaction flow. A: gaze‑pinch selects a single object. B: a light gaze sweep advances selection across neighboring items for rapid multi‑selection. C: an optional voice request specifies the intended operation (for example, ``compare''). D: the system presents a compact, anchored comparison with key attributes. E: aiming a held object toward another establishes a transient pair and expands the context so the system can propose relations for that category pair.}
    \label{fig:interact_flow}
\end{figure}
\noindent

\subsubsection{Selecting objects}
\label{sec:impl-select}
Selection is gaze–directed with pinch to confirm (\autoref{fig:interact_flow}A). Repeated pinches add to the current selection set so that users can nominate multiple objects without leaving the scene. To accelerate multi‑selection, users can sweep their gaze along a row of items and perform a light swipe gesture; the system advances selection to the next plausible neighbor in that direction (\autoref{fig:interact_flow}B). The current selection set becomes the nucleus of the active context: nodes are flagged as ``in focus,'' nearby anchors are temporarily upweighted, and prior selection order is recorded to support procedural inference. The system immediately proposes relations among the selected nodes, and if several types are plausible the confirmation UI described above appears.

\subsubsection{Requesting with voice}
\label{sec:impl-voice}
Users can issue spoken requests that refer to objects deictically (``these two''), by category (``the USB cables''), or by name. Speech is transcribed and resolved against the graph using the current context and scene labels. The request is then routed to inference with the resolved endpoints and desired operation. Voice therefore functions as a high‑level operator on the substrate: it specifies which relation to surface (for example, ``compare'') while the graph supplies the concrete nodes (\autoref{fig:interact_flow}C). Successful responses add or update edges and render edge‑specific overlays, such as the anchored attribute comparison in \autoref{fig:interact_flow}D.

\subsubsection{Gesturing, grabbing, and proximity}
\label{sec:impl-gesture}
Embodied actions also write into the graph. When a user grabs an anchored object, the node enters a held state and follows the hand pose. Aiming the held object toward another item establishes a transient pair; the context expands to include both nodes and the reasoning layer proposes relations that are common for that category pair (\autoref{fig:interact_flow}E). If the user holds two items simultaneously, the system prioritizes comparison and places a compact attribute view on the connector, as illustrated in \autoref{fig:interact_flow}D. As the user moves, nodes within a proximity band around the body are upweighted in the active context, which biases inference toward the user's immediate workspace. Releasing objects clears the held state and decays transient edges unless they have been confirmed by voice or selection.



\add{
\subsection{System Latency and Error Handling}
\label{sec:error_handling}

The system implements timeout safeguards and retry logic for MLLM requests. Failed detections or malformed responses are silently omitted rather than displayed. Users can correct misinterpretations by narrowing their selection or issuing a clarifying voice command (e.g., ``No, connect the cable to the \textit{monitor} port''), triggering re-inference on the updated context.
}

\section{Prototype User Study}
\label{sec:evaluation}

We built the \SystemName and conducted an exploratory user study to examine how inter-object connectivity, physical object manipulation, and contextual sensing affect user performance and experience in AR-supported, object-centric tasks. 
We compared \SystemName{} to a baseline system, \Baseline{}, which we modeled after \textit{XR-Objects}~\cite{doganAugmentedObjectIntelligence2024} and adapted to our HMD-based setting. 
The study employed a within-subjects design with Latin Square counterbalancing across tasks and conditions. Each participant completed three distinct tasks in each condition. The study was conducted using an Apple Vision Pro, though we note that \SystemName{} is designed to operate on any HMD with a front-facing RGB camera.

We also designed a second benchmarking study as a comparative judgment task using short video demonstrations of different scenarios beyond the one participants experienced. The goal of this second part was to characterize when environment-anchored connectivity provides measurable benefit for planning, sequencing, disambiguation, and safety relative to tools that are widely used today. \add{We selected comparators that represent the tools users already reach for when performing physical tasks. \textit{YouTube} is the dominant medium for procedural guidance such as cooking, bike repair, and equipment setup; \textit{ChatGPT} represents the current standard for ad-hoc knowledge retrieval and product comparison; and the \textit{Baseline} (Single-Object AR) acts as a controlled ablation to isolate the specific benefits of connectivity overlays. Not every comparator is equally natural for every scenario (e.g., video tutorials are less typical for shopping), but together they span the breadth of current practice and allow us to characterize where connectivity provides the greatest marginal benefit.}

\subsection{Materials and methods}

A part of our \SystemName prototype we also implemented a baseline method as many people had never interacted with AI or AR, so they had something to compare against. We based our baseline (Single-Object Q\&A) on \textit{XR-Objects}~\cite{doganAugmentedObjectIntelligence2024}, which introduced the concept of AOI and represents the current state of the art in this domain. To enable a fair comparison with \SystemName, we re-implemented the system for the Apple Vision Pro, ensuring both conditions could operate with the same hardware.

\Baseline supports single-object interaction using raycast targeting and standard XR gaze-and-pinch gestures. Users are able to select object anchors and issue verbal queries, which are processed by an MLLM to retrieve object-specific information. In contrast to \SystemName, \Baseline does not support multi-object relationships, physically grounded interaction, or context sensing.

\subsubsection{Participants}
We recruited 12 participants ($n$ = 12) from a local university community (mean age = 24.5 years, SD = 2.2). Participants self-reported their gender as five women and seven men. All participants had normal or corrected-to-normal vision.

\add{
\subsubsection{Tasks}
We designed three task types to stress-test specific aspects of the semantic graph (see \autoref{tab:tasks}):

\begin{itemize}
    \item \textbf{T1 (Compatibility):} Required determining valid edges between a source and multiple targets (e.g., matching cables to ports). This tests the system's ability to externalize \emph{Compatibility} and \emph{Affordance} constraints, replacing mental trial-and-error.
    \item \textbf{T2 (Classification):} Involved sorting a clutter of items based on a hidden property (e.g., recyclability). This evaluates \emph{Similarity} relations, testing if visual grouping reduces search time compared to item-by-item inspection.
    \item \textbf{T3 (Operation):} Required executing a multi-step sequence across physically separated devices (e.g., IoT pairing). This tests \emph{Procedural} and \emph{Causality} links, assessing if spatial routing of instructions improves execution accuracy.
\end{itemize}
}

\begin{table*}[t]
\begin{tabularx}{\textwidth}{@{} l >{\itshape\raggedright\arraybackslash}p{0.22\textwidth} X X @{} }
\toprule
\textbf{ID} & \textbf{Task Type} & \textbf{Task Set A} & \textbf{Task Set B} \\
\midrule
    T1 & Assessing compatibility or connectivity &
    Find out whether Tylenol can be taken with any of these beverages. &
    Find out how to connect the phone to each of these devices. \\
    \addlinespace[0.3em]
    T2 & Classifying objects by properties &
    Identify which of these items might contain gluten. &
    Identify which of these items can be recycled. \\
    \addlinespace[0.3em]
    T3 & Operating or configuring objects &
    Find out how to use the phone to control the IoT LED bulb. &
    Find out how to display distance using the sensor on the Arduino screen. \\
\bottomrule
\end{tabularx}
\caption{Task types and their corresponding instances in \textit{Task Set A} and \textit{Task Set B}.}
\label{tab:tasks}
\end{table*}

\begin{figure}[ht]
    \centering
    \includegraphics[width=\linewidth]{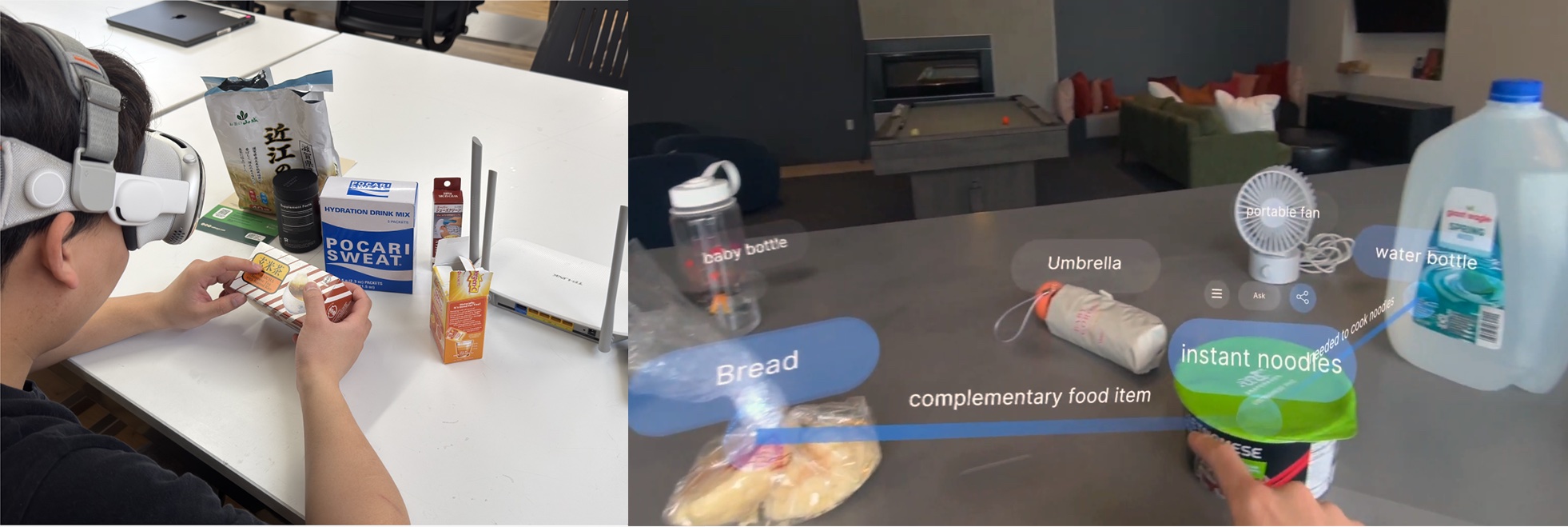}
    \caption{The user study setup included multiple objects for task-based interaction. Participants remained seated, and the objects were rearranged between participants.}
    \label{fig:userstudy}
\end{figure}

Participants completed a pre-study questionnaire, then received an 8-minute training session before beginning with either \SystemName{} or \Baseline{} (counterbalanced). Each condition involved three sub-tasks reflecting common multi-object demands (\autoref{tab:tasks}): assessing compatibility (\textbf{T1}), classifying objects (\textbf{T2}), and operating/configuring objects (\textbf{T3}). Task completion typically took approximately 12 minutes per condition, followed by a post-condition questionnaire. After both conditions, participants completed a semi-structured interview.

\begin{figure}[ht]
    \centering
    \includegraphics[width=.5\linewidth]{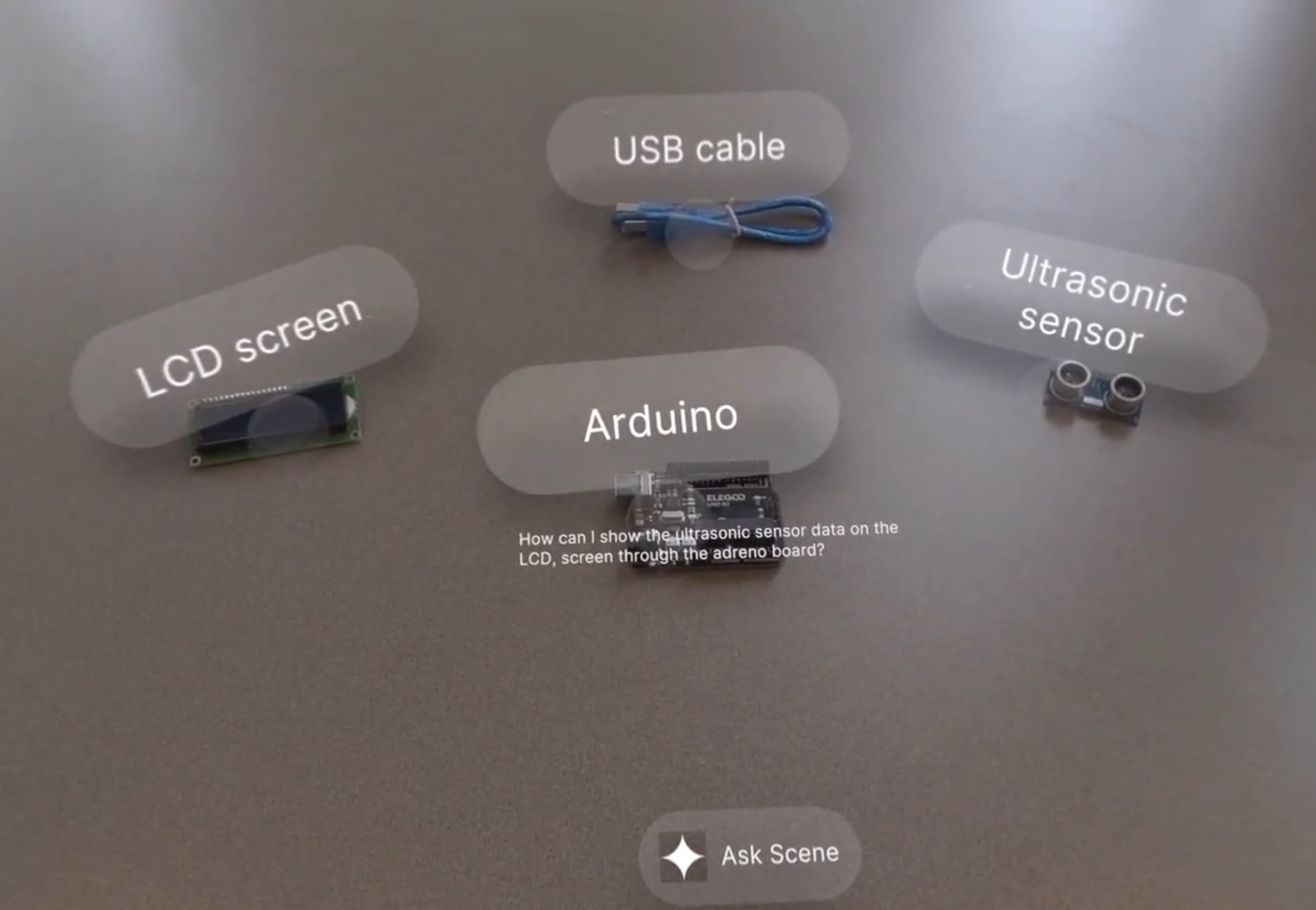}
    \includegraphics[width=.435\linewidth]{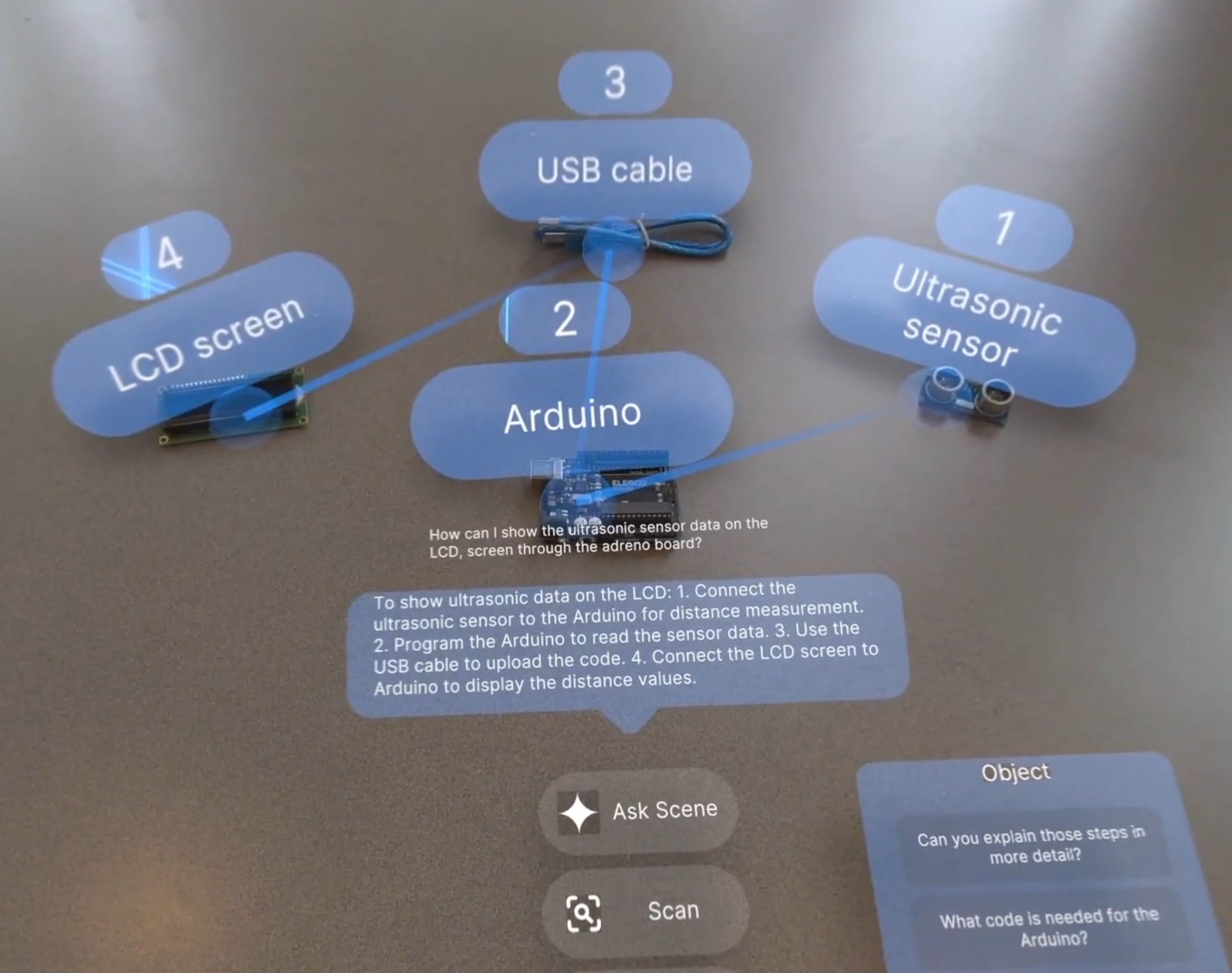}
    \caption{Left: The baseline condition enabled users to interact with the objects identified by the MLLM via verbal queries. Right: In the \SystemName prototype users could visualize the information embedded on the real world seeing via connections.}
    \label{fig:baseline}
\end{figure}

In part 2, participants benchmarked \SystemName against four alternatives via video-based scenarios: real world (unaided), YouTube, Chat Assistant, and \Baseline{}. Five scenarios spanned shopping, cooking, bike maintenance, electronics assembly, and makerspace tasks. For each, participants viewed short clips demonstrating each option, then ranked them and provided Likert ratings on planning support, disambiguation, efficiency, safety confidence, and visual clarity.

\subsubsection{Measures}

Subjective feedback was collected via a post-condition questionnaire adapted from HALIE~\cite{leeEvaluatingHumanLanguageModel2024} (\autoref{tab:post-condition}), NASA TLX for perceived task load, and a semi-structured interview (\autoref{tab:interview}). For the benchmarking survey, the primary outcome was rank order preference per scenario, with secondary Likert ratings on relation-centric criteria.

\subsection{Results}
\label{sec:results}

\begin{figure}[ht]
    \centering
    \includegraphics[width=\linewidth]{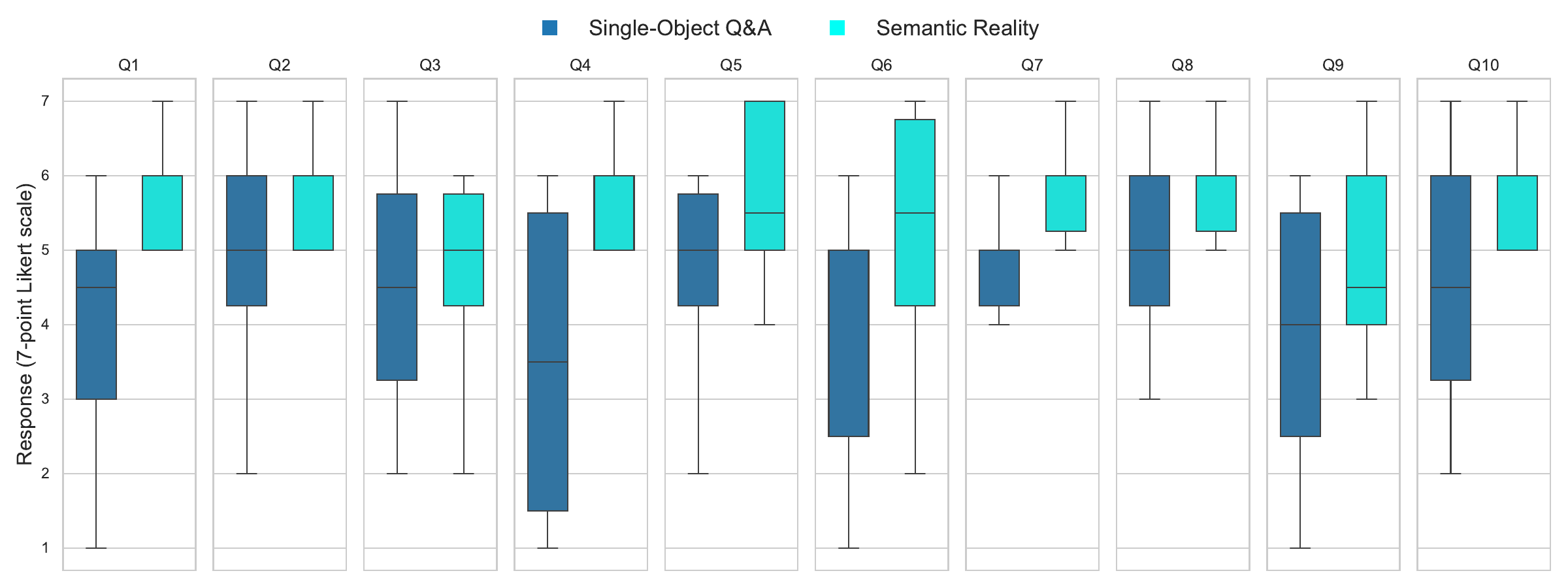}
    \caption{Boxplot of adapted HALIE questionnaire responses on a 7-point Likert scale.}
    \label{fig:qresponses}
\end{figure}

\subsubsection{HALIE Questionnaire}

Wilcoxon tests showed significant preferences for \SystemName on multiple items (\autoref{fig:qresponses}): helpfulness (Q1, $M_{SR}=5.25$ vs. $M_{BL}=4.08$, $p=.015$), inter-object clarity (Q4, $M_{SR}=5.5$ vs. $M_{BL}=3.3$, $p=.016$), engagement (Q7, $p=.047$), and satisfaction (Q8, $p=.047$). Context understanding (Q5) showed a trend favoring \SystemName ($p=.05$). Control items such as responsiveness (Q3, $p=.86$) showed no difference, confirming that results were not driven by backend performance differences, as both systems used the same AI engine.

\subsubsection{NASA TLX}

\begin{figure}[ht]
    \centering
    \includegraphics[width=\linewidth]{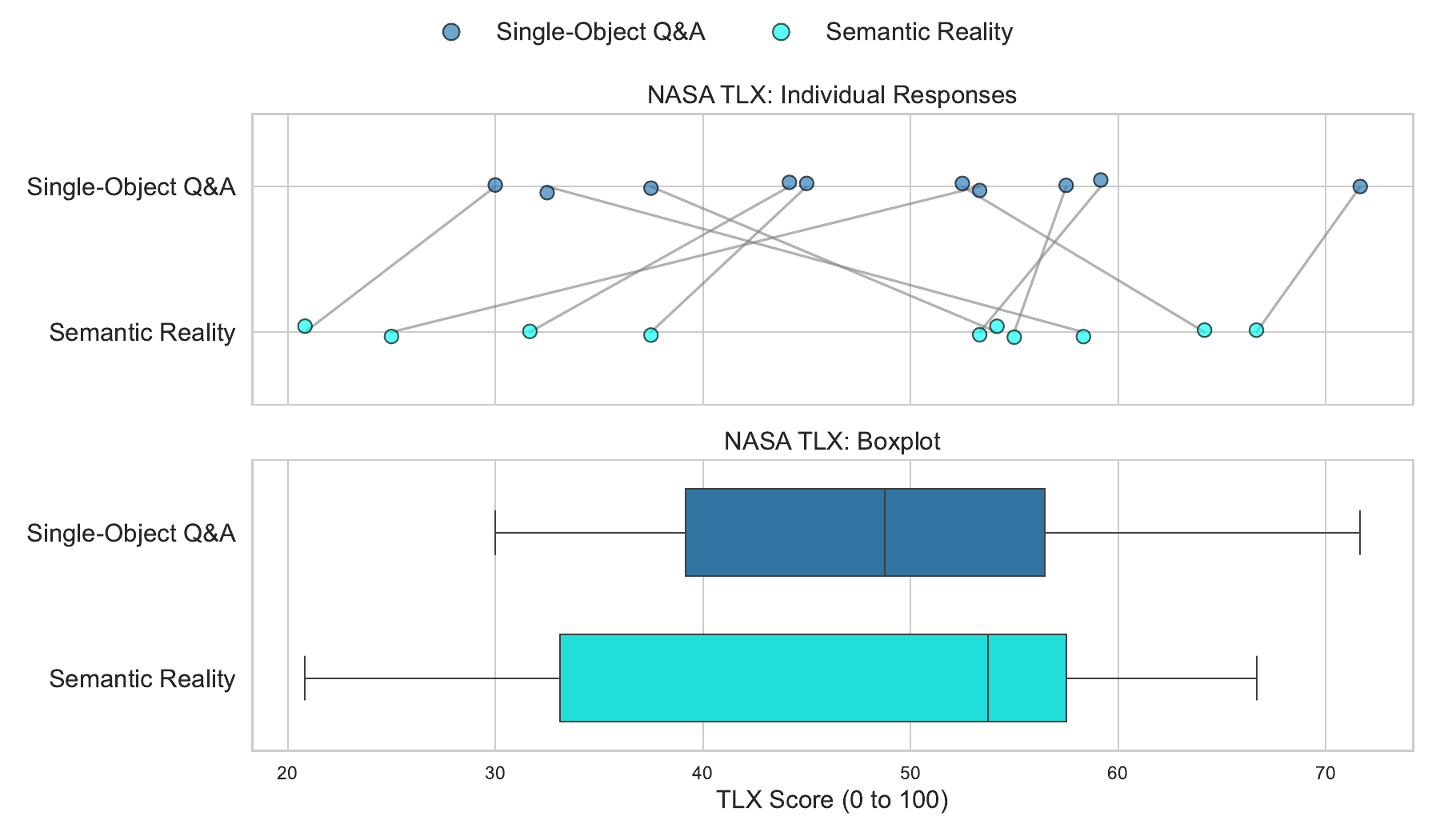}
    \caption{NASA TLX questionnaire responses.}
    \label{fig:nasatlx}
\end{figure}

\noindent
Overall perceived task load was lower for \SystemName ($M=40.4$, $SD=22.2$) than Baseline ($M=49.4$, $SD=15.1$), though not significant ($Z=1.53, p=0.13$). Mental Demand and Frustration showed similar non-significant trends favoring \SystemName. Most notably, participants reported significantly better \textbf{Performance} with \SystemName ($M=33.8$, $SD=21.3$) versus Baseline ($M=55.8$, $SD=29.1$; $Z=2.27, p=0.02$), indicating that despite more visual information, users felt more successful.

\add{\subsubsection{Task Completion Time}
Participants spent $86.3$s ($SD=26.7$) per task with \SystemName versus $77.8$s ($SD=22.1$) with Baseline, a non-significant difference ($Z=0.82, p=0.41$), suggesting the additional visual information did not introduce substantial overhead.}

\subsubsection{Semi-Structured Interview}
Three themes emerged from the interview transcripts.

\paragraph{\textbf{Inter-object connectivity supported multi-object reasoning.}}
Participants highlighted the system's ability to expose relationships and support multi-object workflows. P10 noted that ``when I track an object, it has a line that connects to other objects,'' and P8 found it ``really explains [connections] in a really good manner'' during assembly tasks. Several noted it was ``more helpful in a more complex task engaging more objects'' (P3), while the baseline felt limited in handling relationships (P4, P10).

\paragraph{\textbf{Context-aware suggestions reduced effort.}}
Participants found that \SystemName's ability to infer relevant connections improved task alignment. P10 appreciated that ``it just comes up with the idea of what I'm going to do,'' and P7 found that the system ``knows what I want to know from the context.'' P6 noted ``I don't have to click each of them,'' while \Baseline{} lacked environmental context (P4).

\paragraph{\textbf{Situated assistance for everyday environments.}}
Participants envisioned using the system in supermarkets (P2), for recycling decisions (P1), and during cooking (P7). The ability to visually link related objects and surface relevant distinctions was particularly valued for tasks involving implicit structure and multi-item comparisons (P11).

\label{sec:benchmarking}

\subsubsection{Benchmarking survey}
We analyzed per-scenario rank preferences with Friedman tests (Kendall's $W$) and planned pairwise comparisons (\SystemName vs. \Baseline/YouTube/Chat) using Wilcoxon signed-rank tests with Holm correction. For HALIE outcomes, we formed a per-participant composite (Cronbach's $\alpha$ verified) and ran Friedman tests across modalities with Holm-corrected Wilcoxon pairs.

\begin{figure*}[h]
  \centering
  \includegraphics[width=.48\linewidth]{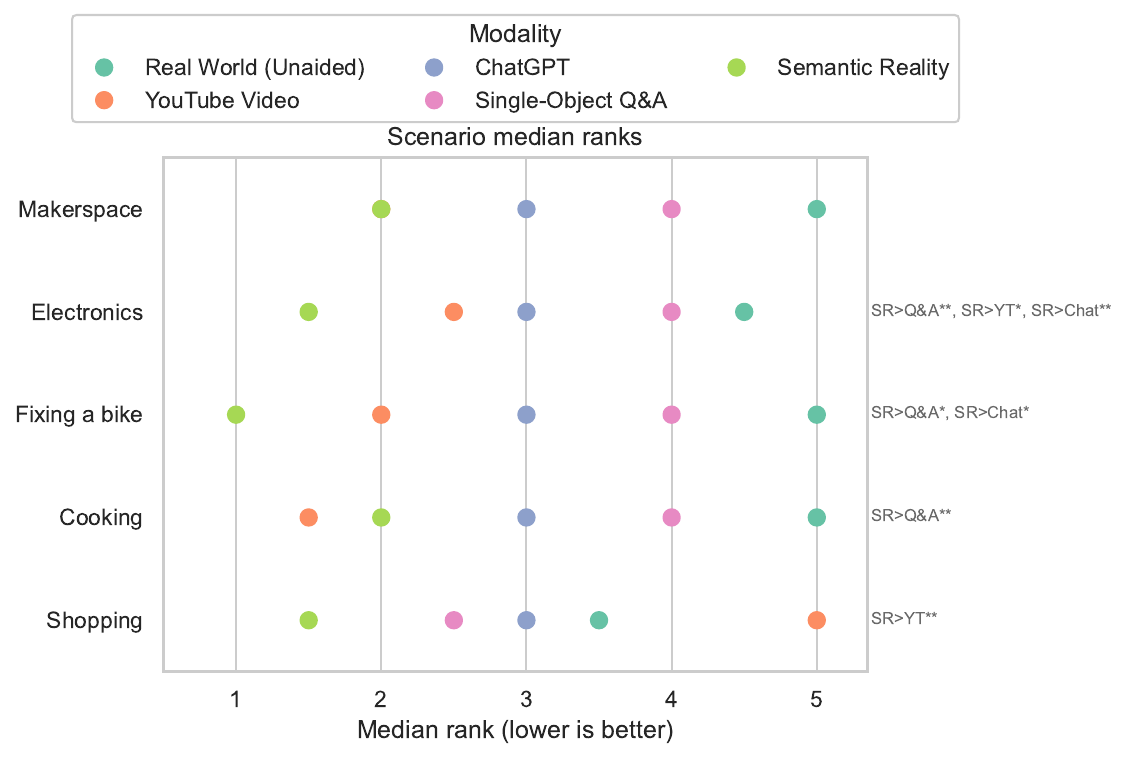}
  \hfill
  \includegraphics[width=.48\linewidth]{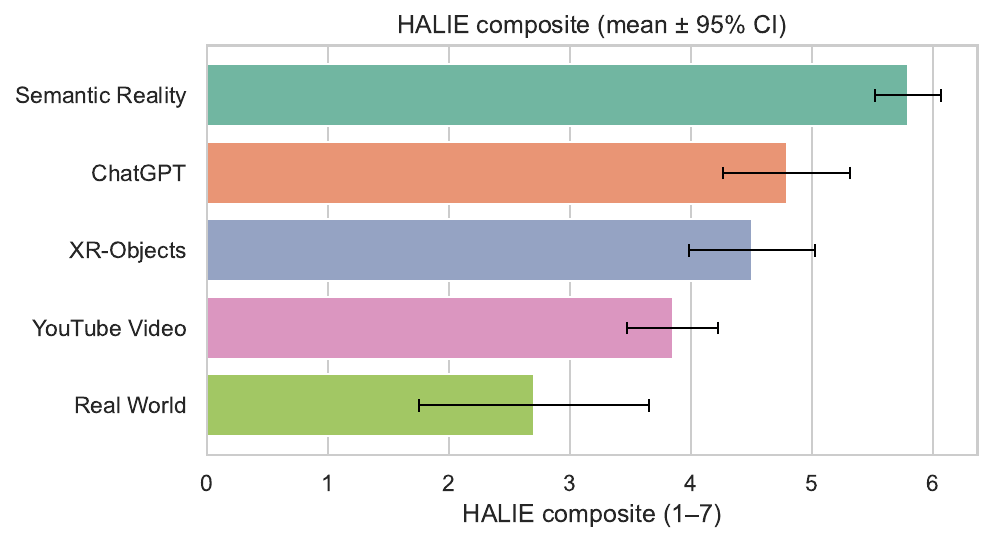}
  \caption{Left: Scenario median ranks (lower is better) with Holm-corrected significant pairwise wins for \SystemName annotated per scenario. Right: HALIE composite (mean \textpm{} 95\% CI) by modality.}
  \label{fig:benchmarking-ranks}
\end{figure*}

\textbf{Scenario preferences.} \SystemName achieved the best or tied-best median rank in 4/5 scenarios (Shopping 1.5, Cooking 2.0, Bike 1.0, Electronics 1.5, Makerspace 2.0; lower is better). Pairwise Wilcoxon tests (Holm-corrected within scenario) showed that \SystemName significantly outperformed \Baseline{} in Cooking ($p_{\text{holm}}=.009$), Bike ($p_{\text{holm}}=.016$), and Electronics ($p_{\text{holm}}=.003$). \SystemName also outperformed YouTube in Shopping ($p_{\text{holm}}=.006$) and Electronics ($p_{\text{holm}}=.047$), and outperformed Chat in Bike ($p_{\text{holm}}=.020$) and Electronics ($p_{\text{holm}}=.0015$). Differences were not significant in Makerspace (all $p_{\text{holm}}\ge .23$), and YouTube was strongest in Cooking (YouTube median 1.5).

\textbf{HALIE composite.} The 10-item HALIE scale exhibited acceptable reliability across modalities (Cronbach's $\alpha=0.68$--$0.95$). The composite differed by modality (Friedman $\chi^2=27.65$, $p=1.47\times 10^{-5}$, $W=0.576$). Planned pairs (Holm-corrected across three comparisons) favored \SystemName over \Baseline{} (mean diff $+1.28$, 95\% CI $[0.91, 1.65]$, $p_{\text{holm}}=.0015$), over YouTube ($+1.94$, $[1.49, 2.38]$, $p_{\text{holm}}=.0015$), and over Chat ($+1.00$, $[0.48, 1.53]$, $p_{\text{holm}}=.0063$).

\begin{figure}[h]
  \centering
  \includegraphics[width=\linewidth]{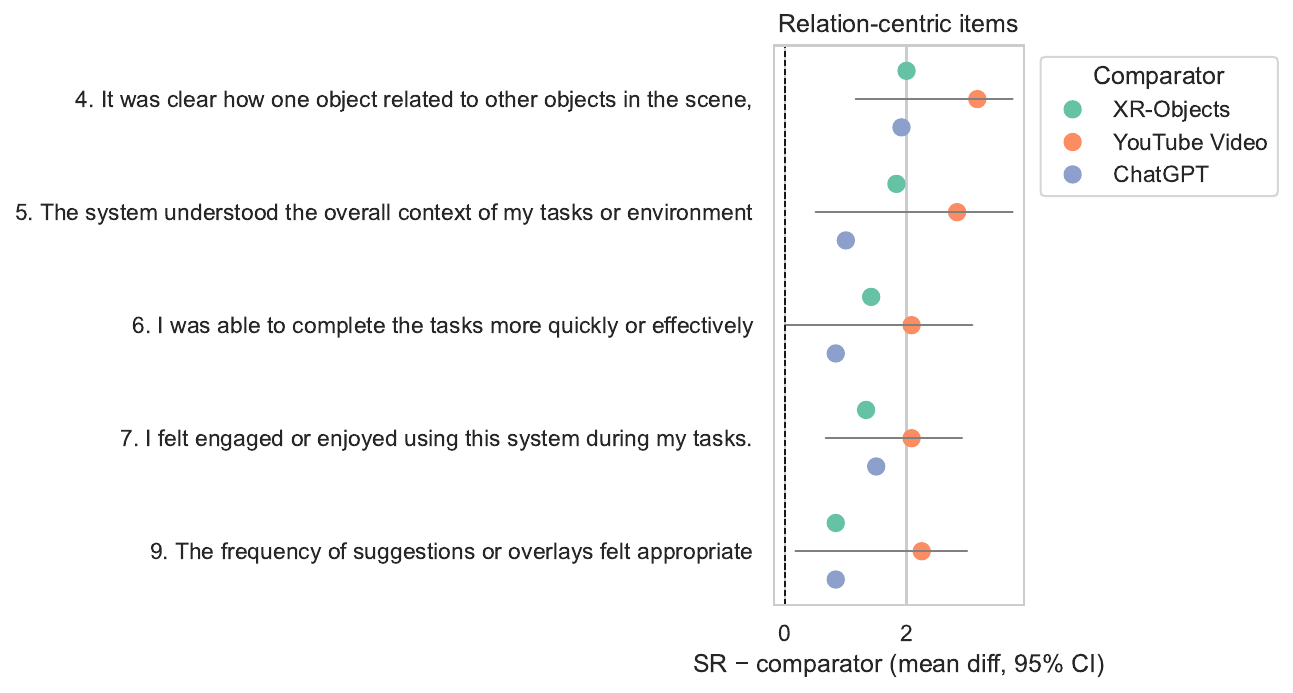}
  \caption{Relation-centric items: SR minus comparator (mean difference \textpm{} 95\% CI).}
  \label{fig:benchmarking-items}
\end{figure}

\textbf{Item-level patterns.} Gains concentrated on relation-centric outcomes. For \emph{relation clarity} (``how one object related to others''): \SystemName $>$ \Baseline{} ($\Delta=+2.00$, 95\% CI $[1.17, 2.75]$, $p_{\text{holm}}=.029$), $>$ YouTube ($\Delta=+3.17$, $[2.50, 3.75]$, $p_{\text{holm}}=.0049$), and $>$ Chat ($\Delta=+1.92$, $[1.17, 2.75]$, $p_{\text{holm}}=.0195$). For \emph{context understood}: \SystemName $>$ \Baseline{} ($\Delta=+1.83$, $[1.08, 2.50]$, $p_{\text{holm}}=.035$) and $>$ YouTube ($\Delta=+2.83$, $[2.00, 3.75]$, $p_{\text{holm}}=.0049$). For \emph{efficiency}: \SystemName $>$ \Baseline{} ($\Delta=+1.42$, $[0.75, 2.00]$, $p_{\text{holm}}=.035$) and $>$ YouTube ($\Delta=+2.08$, $[1.17, 3.08]$, $p_{\text{holm}}=.024$). We also observed higher \emph{engagement} (vs. \Baseline{} $\Delta=+1.33$, $p_{\text{holm}}=.035$; vs. YouTube $\Delta=+2.08$, $p_{\text{holm}}=.015$) and appropriate \emph{suggestion frequency} (vs. \Baseline{} $\Delta=+0.83$, $p_{\text{holm}}=.035$; vs. YouTube $\Delta=+2.25$, $p_{\text{holm}}=.0078$). Ease-of-use differences were small and not consistently significant across pairs.


The benchmarking results indicate that environment-anchored connectivity is particularly beneficial when tasks involve multi-object coordination, ordering, or compatibility constraints. In these cases, visualizing relationships in situ, together with physical manipulation, appears to reduce the cost of mental indexing across objects and steps. In contrast, when the task is primarily passive learning or single-object lookup, YouTube or a chat assistant can be competitive.


\section{Discussion}
We have presented \SystemName, a system that expands AR+AI interactions from single-object queries to a connectivity-centered substrate. Our evaluation demonstrates that this approach improves perceived performance and engagement without significantly increasing perceived task load. In this section, we discuss the trade-off between visual complexity and cognitive utility, the role of the connectivity substrate in grounding AI, and current limitations.

\add{\subsection{The Cost of Connectivity: Visual vs. Cognitive Load}}
\add{A key concern in AR interface design is that adding information, specifically rendering multiple connection lines and labels, will inevitably increase cognitive load and visual clutter. Our results offer a more nuanced view. While \SystemName introduces more visual elements than the baseline, participants did not report higher Mental Demand ($M_{SR}=51.7$ vs. $M_{BL}=55.4$) or Effort ($M_{SR}=43.8$ vs. $M_{BL}=47.5$). Crucially, they reported significantly better perceived Performance ($p=0.015$).}

\add{This pattern is consistent with a ``cognitive offloading'' hypothesis: by externalizing relationships into the world (e.g., showing exactly which cable fits where), the system may reduce the \emph{intrinsic} cognitive load required to maintain a mental model of the task, offsetting the added \emph{extraneous} visual load. One participant (P9) explicitly contrasted the systems, noting that the baseline felt like ``a camera taking a picture,'' whereas \SystemName felt ``more like human brain thinking'' because it allowed them to visually ``pick what you want to know'' from a web of relations rather than holding the state in memory. However, our controlled setting and relatively simple tasks limit the generalizability of this finding; more complex environments with greater object density may shift this balance.

In practice, the context window and intent classifier constrain each interaction to one or two relation types at a time; situations in which many concurrent edge types appear were rare during our study. Nevertheless, our video scenarios and qualitative feedback indicate that clutter becomes a risk when low-relevance relations (e.g., spatial links) accumulate. Future systems should therefore treat connectivity as a ``progressive disclosure'' layer, showing only the most relevant links by default and allowing users to explicitly request broader context (e.g., ``show me all comparisons'') when needed.}

\add{\subsection{Grounding AI in a Shared Substrate}}
\add{By mediating interactions through a persistent semantic graph, \SystemName bridges the gap between ``chatbot'' AI and physically embodied agents. Participants naturally used deictic language (e.g., ``these two''), relying on visual selection to disambiguate intent. Unlike purely generative approaches, the substrate enforces a schema (the 8 relation types) that constrains outputs to actionable, verifiable links. When a link is wrong, it can be rejected as a discrete edge, unlike errors buried in immutable model output.}

\add{\subsection{Managing Visual Complexity}
\label{sec:visual_complexity}
As the number of objects and relations grows, visual clutter can become a challenge. Our prototype uses ephemeral edges and depth testing, but a scalable system must incorporate advanced view management techniques~\cite{bell2001view, kalkofen2008comprehensible, veas2012extended}. Key strategies include progressive disclosure (showing only the most salient relations by default), user-controlled filtering by relation type, context-aware fading of relations outside the user's immediate workspace, and automated layout to minimize edge crossings and occlusion. Integrating these strategies is a critical next step.}

\subsection{Limitations}
Several factors constrained our current prototype and study. First, \SystemName depends on real-time mesh reconstruction to anchor objects and sub-regions. In dynamic settings or with fast-moving objects, participants occasionally encountered drift, which sometimes broke the illusion of stable object references. Addressing this may require improved on-device 3D tracking or more frequent re-anchoring triggered by user interactions. Second, because the system relies on external MLLM queries, response times can slow if many requests are issued concurrently. We expect next-generation hardware, possibly running smaller but specialized vision-language models on-device, to reduce these delays and support richer concurrency.

\add{Finally, our evaluation has limitations regarding validity and generalizability. First, the study was conducted in a controlled laboratory setting with a curated set of objects to ensure consistent tracking. Real-world environments often present greater visual clutter, variable lighting, and unpredictable arrangements that may challenge both the perception pipeline and the user's ability to parse connectivity overlays. Second, while we selected tasks and comparators (e.g., YouTube, Chat) to span common scenarios, they do not capture the full diversity of physical workflows, which may limit the external validity of our benchmarking results. Third, our sample size ($N=12$) and demographics (university population) are relatively small and homogeneous. Future long-term deployments in diverse contexts with broader user groups are necessary to understand how \SystemName adapts to the messiness and variety of everyday life.}

\section{Conclusion}
We introduced \SystemName, a connectivity-centered substrate for AR+AI that represents scenes as a dynamic semantic graph of objects and typed relations. By organizing connectivity through eight relation types, interaction initiative, and an agent-centric context window, the system enables in-situ, referential guidance for planning, comparison, and assembly.
Our prototype couples open-vocabulary perception with constrained reasoning to infer and render relations as consistent overlays. In an exploratory study and a complementary benchmarking comparison, participants reported clearer inter-object understanding and higher engagement without increased task load. Future work will scale to larger scenes, add richer correction and provenance cues, and integrate more deeply with device and IoT capabilities.



\bibliographystyle{ACM-Reference-Format}
\bibliography{bib/zotero,bib/manual}

\appendix

\add{
\section{System Prompts}
\label{appendix:prompts}

To facilitate reproducibility, we provide the core system prompts used to drive the MLLM (Gemini 2.5 Flash) for object detection, relationship inference, and task planning. Note that some implementation-specific formatting (e.g., JSON schema definitions) has been simplified for readability.

\noindent\textbf{Object Detection.} Detect distinct MAIN OBJECTS in this image. Focus on independent, standalone objects rather than small details, labels, or parts of objects. For each main object, return a JSON object with a 2D bounding box, a concise label (max 4 words), and a descriptive sentence. \textbf{Constraints:} Focus on main entities only; avoid small details like ``label'', ``sticker'', ``button''; use concrete identifiers (Brand + Object Type).

\noindent\textbf{Voice Command Planning.} You are an AR assistant planning a visualization. \textbf{User Request:} [Transcribed Text]. \textbf{Available Objects:} [List of scene objects]. \textbf{Task:} Select a relation type from the 8 supported types and choose the minimal set of objects needed to satisfy the user's request.

\noindent\textbf{Relationship Inference.} The following prompts (\autoref{tab:prompts}) are used to infer specific edge types between selected nodes.

\begin{table}[H]
\caption{System prompts used for relationship inference.}
\label{tab:prompts}
\small
\begin{tabularx}{\linewidth}{l X}
\toprule
\textbf{Relation Type} & \textbf{Prompt Details} \\
\midrule
\textbf{Type Selection} & \textbf{Prompt:} Analyze the relationship between these objects and determine the SINGLE BEST relationship type that applies. \newline \textbf{Input:} List of selected objects. \newline \textbf{Available Types:} spatial, structural, similarity, comparison, affordance, compatibility, procedural, causality. \newline \textbf{Output:} JSON with the chosen type, confidence (0.0-1.0), and reason. \\ \addlinespace
\textbf{Comparison} & \textbf{Prompt:} Compare these two objects: [Object A] and [Object B]. You MUST provide exactly 3 attributes that differ between these objects, focusing ONLY on practical consumer benefits. \newline \textbf{Constraints:} Forbidden (Brand names, visual differences); Focus (Cost, ease of use, value). \\ \addlinespace
\textbf{Similarity} & \textbf{Prompt:} The user selected several physical objects in an AR scene. Determine if they belong to the same general type or share a strong common theme. \newline \textbf{Output:} JSON with boolean `sameType`, short `theme` title, and summary. \\ \addlinespace
\textbf{Structural} & \textbf{Prompt:} Detect if any selected items are structural sub-components of another selected item and provide step-by-step assembly instructions. Only consider relationships where every child truly belongs to or is a detachable piece of the parent object. \\ \addlinespace
\textbf{Affordance} & \textbf{Prompt:} Identify tool-object relationships where tools can act upon target objects. Find relationships where one object is a tool that can perform actions on other selected objects (e.g., knife acts-on garlic). \newline \textbf{Output:} JSON with `tool`, `targets`, `action` (e.g., ``crush then chop''), and `tip`. \\ \addlinespace
\textbf{Compatibility} & \textbf{Prompt:} You are a careful safety assistant. Determine if combining or using these two objects together is unsafe, hazardous, or fundamentally incompatible. Consider physical danger, functional incompatibility, and common-sense hazards. \newline \textbf{Output:} JSON with `incompatible` (boolean) and warning. \\ \addlinespace
\textbf{Procedural} & \textbf{Prompt:} Identify the most likely task or procedure that someone would want to do with these objects together. Create detailed, actionable step-by-step instructions. \newline \textbf{Output:} JSON with task name, description, and ordered list of steps. \\ \addlinespace
\textbf{Causality} & \textbf{Prompt:} Analyze causality relationships where one object (cause) produces effects on other objects (e.g., switch causes light to turn on). \newline \textbf{Output:} JSON identifying `cause`, `effects`, `action`, and `consequence`. \\
\bottomrule
\end{tabularx}
\end{table}
}

\section{Evaluation Instruments}
\label{appendix:instruments}

\begin{table}[H]
    \begin{tabular}{@{}p{\dimexpr\linewidth-2\tabcolsep}@{}}
        \toprule
        1. I found this system \textbf{helpful} for completing my tasks. \\ \addlinespace[0.3em]
        2. It was \textbf{easy to learn and interact} with this system, requiring little effort on my part. \\ \addlinespace[0.3em]
        3. The system responded \textbf{promptly and accurately} to my selections or queries. \\ \addlinespace[0.3em]
        4. It was clear how \textbf{one object related to other objects} in the scene, making multi-object tasks easier. \\ \addlinespace[0.3em]
        5. The system \textbf{understood the overall context} of my tasks or environment and offered relevant suggestions or information. \\ \addlinespace[0.3em]
        6. I was able to \textbf{complete the tasks more quickly or effectively} using this system than I would have otherwise. \\ \addlinespace[0.3em]
        7. I felt \textbf{engaged or enjoyed} using this system during my tasks. \\ \addlinespace[0.3em]
        8. I am \textbf{satisfied with how} this system provided answers or instructions for my tasks. \\ \addlinespace[0.3em]
        9. The \textbf{frequency of suggestions} or overlays felt appropriate (neither too few nor too many). \\ \addlinespace[0.3em]
        10. I could see myself using this system for \textbf{real-world, everyday scenarios} if it were available. \\ \addlinespace[0.3em]
        \bottomrule
    \end{tabular}
    \caption{HALIE Questionnaire Statements, Adapted from \cite{leeEvaluatingHumanLanguageModel2024}}
    \label{tab:post-condition}
\end{table}

\begin{table}[H]
\begin{tabular}{@{}p{0.97\linewidth}@{}}
\toprule
1. You tried two different systems. What stood out to you about each one? Were there any specific features or behaviors that you particularly liked or disliked? \\
\addlinespace[0.3em]
2. How did the two systems compare in supporting the tasks you were doing? Were there differences in how easy or natural each one felt to use? \\
\addlinespace[0.3em]
3. One of the systems supported interacting with groups of related objects and their relationships. How did you experience these environment-level interactions? Can you give examples? \\
\addlinespace[0.3em]
4. In one system, your physical actions--like pointing, picking up, or arranging objects--affected how it responded. How did you experience these different types of actions? \\
\addlinespace[0.3em]
5. Did either system feel aware of what you were doing or what was around you? Can you think of any examples where it responded in a way that matched--or did not match--your context? \\
\addlinespace[0.3em]
6. If you could improve one or both systems, what would you change or add? Are there any physical actions or interactions you wish the system had supported? \\
\addlinespace[0.3em]
7. Can you think of situations in your daily life--whether everyday or occasional--where a system like this might or might not be useful?\\
\bottomrule
\end{tabular}
\caption{Semi-structured interview.}
\label{tab:interview}
\end{table}

\section{Relation Example Figures}
\label{appendix:relation_figures}

\begin{figure*}[h!]
    \centering
    \includegraphics[width=\linewidth]{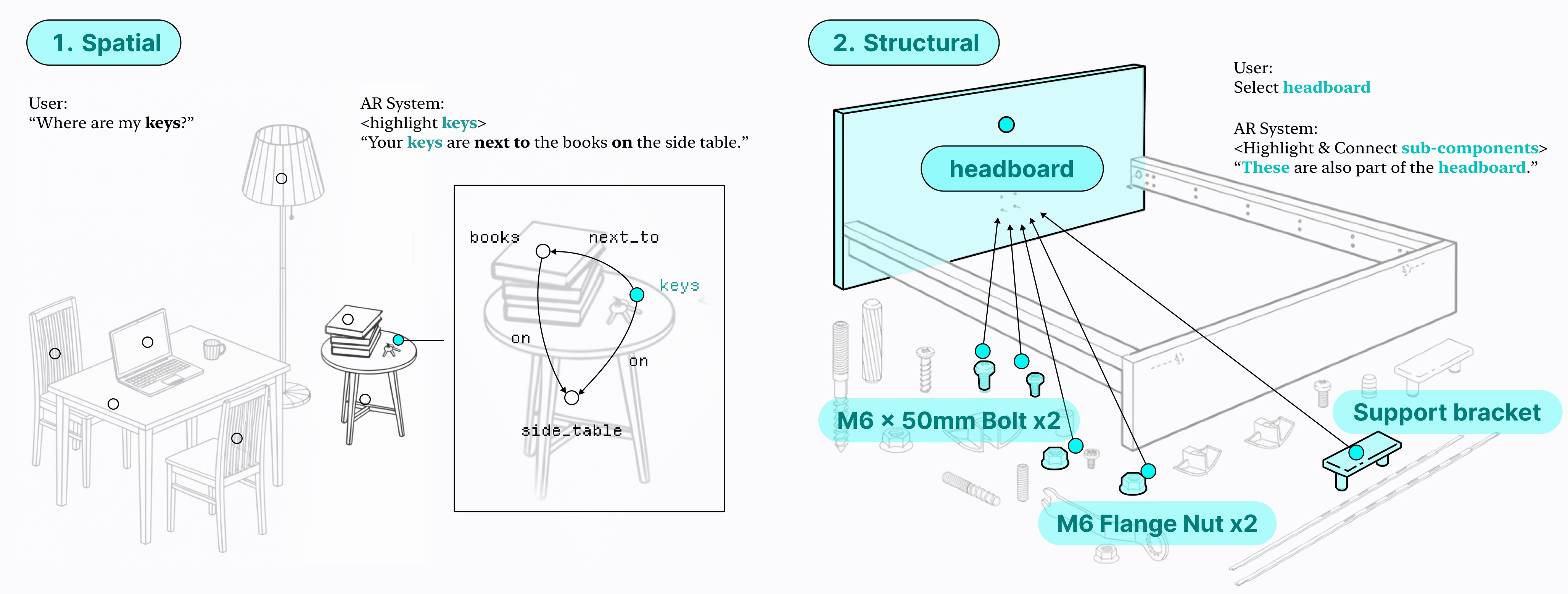}
    \caption{Relation examples (1--2). Left: Spatial: the system localizes an item by describing its position relative to nearby anchors (e.g., keys next to the books on the side table). Right: Structural: selecting a parent (headboard) highlights subcomponents and fasteners as \emph{part-of} the parent (bolts, flange nuts, support bracket).}
    \label{fig:rel_banner_1}
\end{figure*}

\begin{figure*}[h!]
    \centering
    \includegraphics[width=\linewidth]{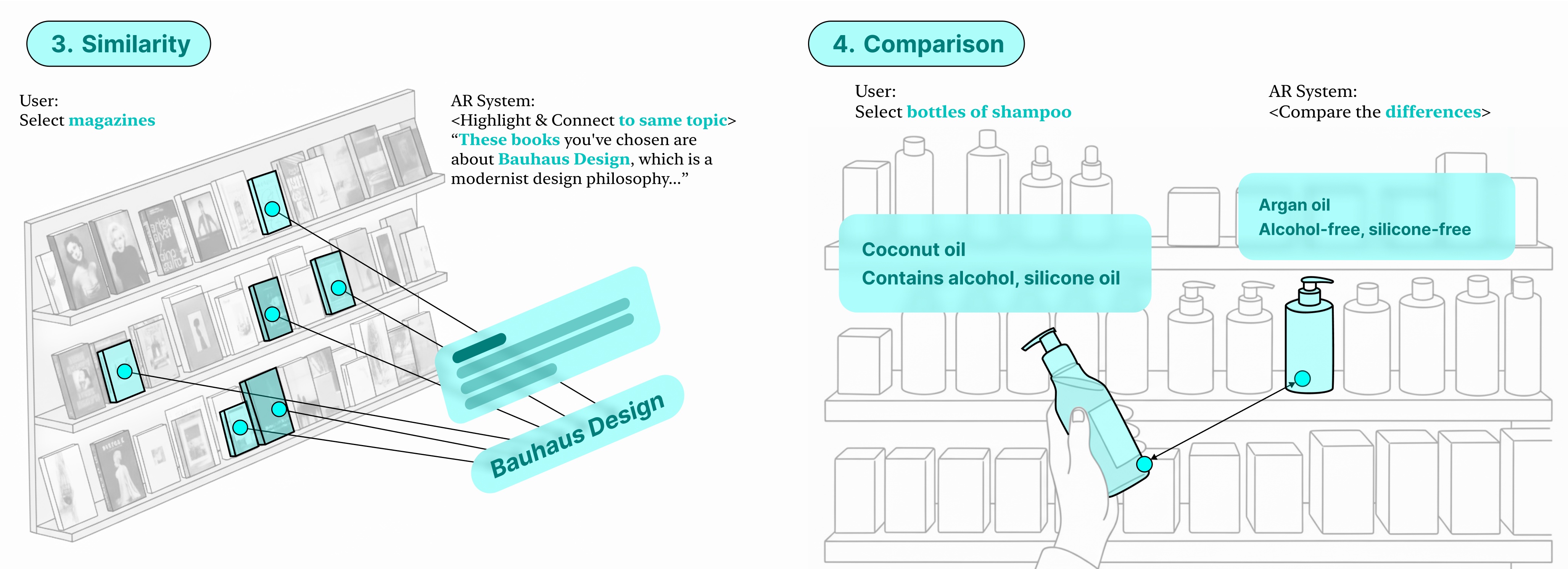}
    \caption{Relation examples (3--4). Left: Similarity: selecting several books with related content surfaces a shared topic label (e.g., ``Bauhaus Design'') and a concise summary, visually clustering the items. Right: Comparison: selecting two shampoos shows side-by-side attributes and highlights differences (e.g., coconut vs argan oil; alcohol-free, silicone-free).}
    \label{fig:rel_banner_2}
\end{figure*}

\begin{figure*}[h!]
    \centering
    \includegraphics[width=\linewidth]{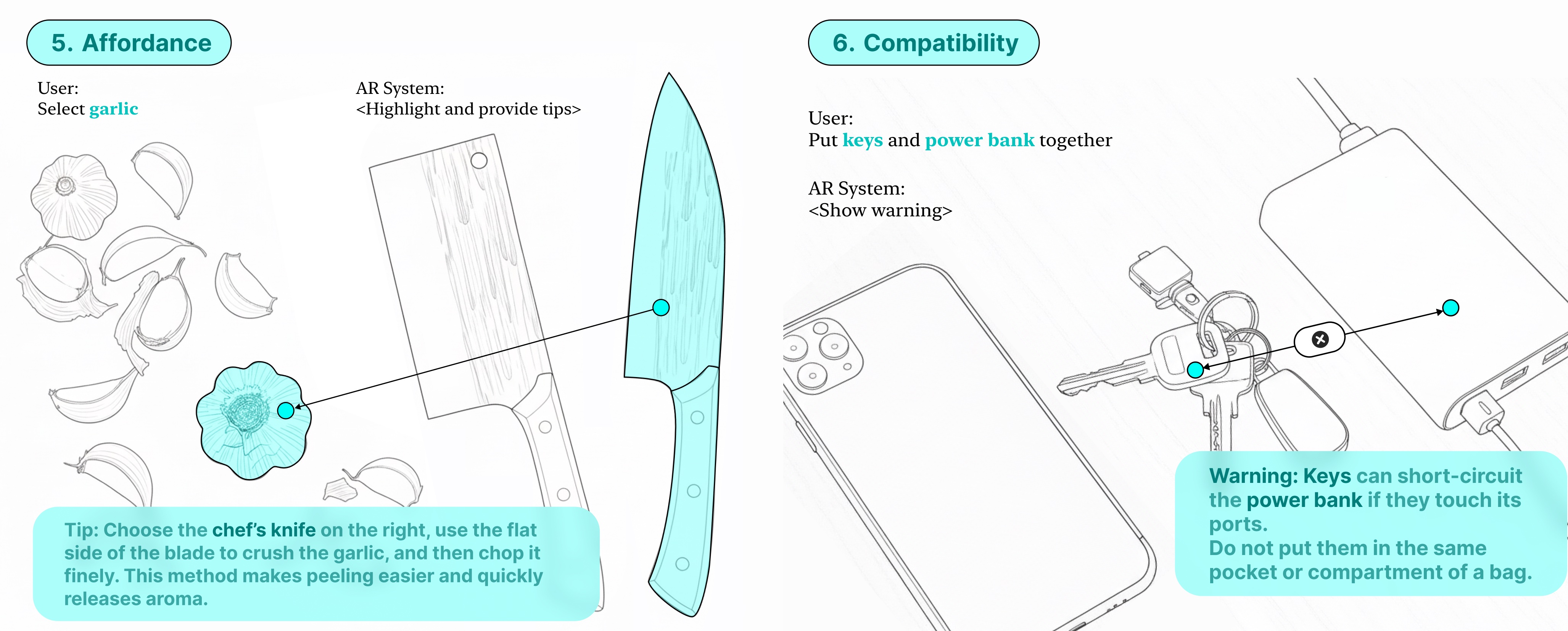}
    \caption{Relation examples (5--6). Left: Affordance: selecting an ingredient (garlic) highlights a suitable tool (chef's knife) and provides a concise technique tip (crush then chop). Right: Compatibility: pairing items prompts a fits-with or safety check; compatible pairs are confirmed with key parameters, while risky or incompatible pairs trigger warnings or alternatives.}
    \label{fig:rel_banner_3}
\end{figure*}

\begin{figure*}[h!]
    \centering
    \includegraphics[width=\linewidth]{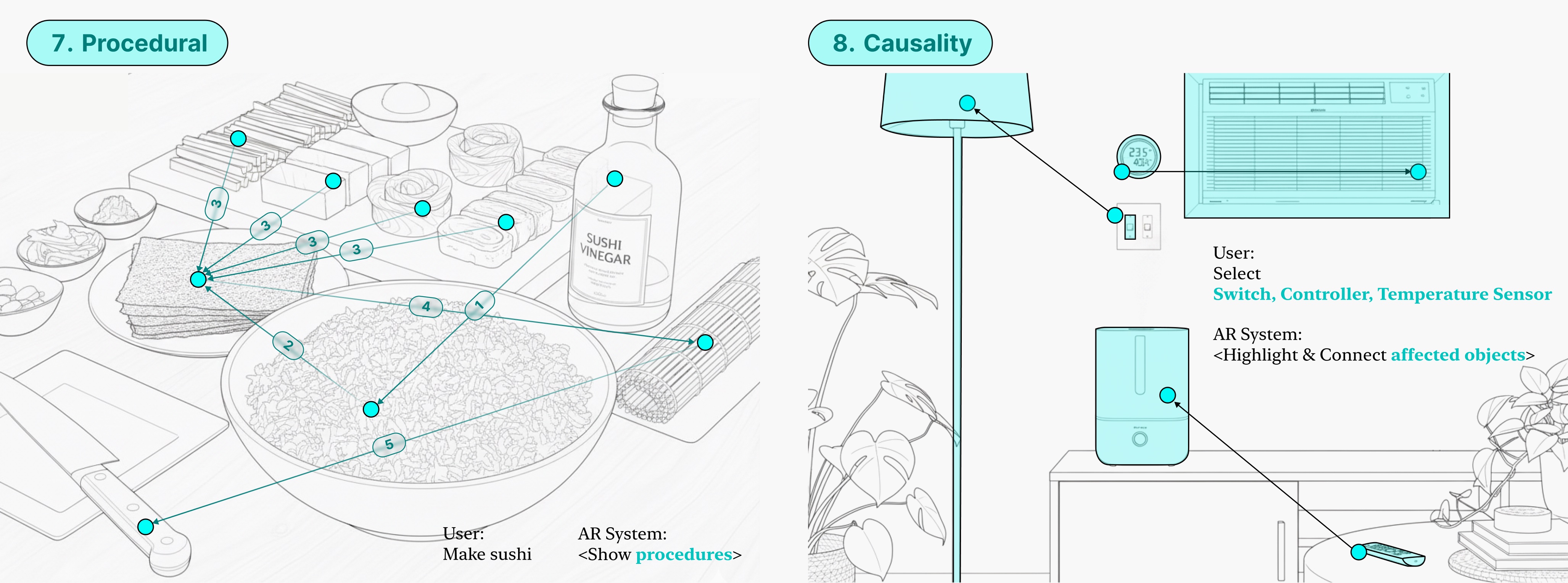}
    \caption{Relation examples (7--8). Left: Procedural: a lightweight plan shows numbered steps and clusters for parallelizable actions, updating as steps complete. Right: Causality: precondition--effect links visualize how actions produce outcomes and allow in-place previews to support what-if reasoning.}
    \label{fig:rel_banner_4}
\end{figure*}

\end{document}